\documentclass[11pt]{article}
\usepackage[utf8]{inputenc}
\usepackage{amssymb,amsmath,mathrsfs,enumerate}
\usepackage{graphicx,rotate,multicol}
\usepackage{float}
\usepackage{tocloft}
\usepackage{subfig}
\usepackage[margin=10pt,labelfont=bf]{caption}
\usepackage{cite}
\usepackage{soul}
\usepackage[colorlinks=true,
            linkcolor=blue,
            urlcolor=blue,
            citecolor=red]{hyperref}

\makeatletter
\let\Hy@linktoc\Hy@linktoc@page
\makeatother

\usepackage{color}
\definecolor{ourcolor}{rgb}{0.7, 0.25, 0.05}

\let\tilde=\widetilde

\let\bar=\overline

\def \order(#1){{\mathcal O} \left(#1 \right)}

\evensidemargin=\oddsidemargin
\allowdisplaybreaks
\title{\color{black}{\bf $t\to cg$ and $t\to cZ$ in Universal Extra Dimensional Models}}

\author {\bf   Cheng-Wei Chiang,$^{a,b}$\footnote{chengwei@phys.ntu.edu.tw} 
\hspace{4pt}   Ujjal Kumar Dey,$^{a,}$\footnote{ujjal@phys.ntu.edu.tw} 
\hspace{4pt}   Tapoja Jha$^{d,e}$\footnote{tapoja.j@iopb.res.in} \\[10pt]
\small\em $^a$Department of Physics, 
		      National Taiwan University,
		      Taipei 10617, Taiwan, R.O.C. \\
\small\em $^b$Institute of Physics, 
			  Academia Sinica, 
			  Taipei 11529, Taiwan, R.O.C. \\
\small\em $^d$Institute of Physics, 
              Sachivalaya Marg, 
              Bhubaneswar, Odisha 751005, India\\
\small\em $^e$Homi Bhabha National Institute,
              Training School Complex, Anushakti Nagar, 
              Mumbai 400085, India
}

\date{}

\begin{document}

\maketitle
                         
\begin{abstract}
In this work we perform a complete one-loop calculation of the flavor-changing top quark decays $t\to cg$ and $t\to cZ$ in the universal extra dimensional models. We find that the branching ratios of these decays in the minimal scenario remain unaltered from the Standard Model expectations for currently allowed values of the inverse compactification radius. In the non-minimal setup, the branching ratios can be enhanced from the Standard Model due to the presence of boundary localized terms which modify the mass spectrum and couplings in the theory in a non-trivial way. We also check the compatibility of the parameter choices that result in enhancements in these rare decays with other experimental observations.
\end{abstract}


\newpage

\hrule \hrule
\tableofcontents
\vskip 10pt
\hrule \hrule

\section{Introduction}

Extra-dimensional models provide an interesting extension of the Standard Model (SM) of particle physics to address some of the unresolved issues within the framework of SM. The possibility of the existence of an extra dimension at the scale of TeV was first put forward in~\cite{Antoniadis:1990ew}, followed by a number of exquisite extra-dimensional models proposed to address various shortcomings of the SM~\cite{ArkaniHamed:1998rs, Antoniadis:1998ig, Randall:1999ee, Randall:1999vf, Appelquist:2000nn}. In this paper, we will consider one of the simplest models of this ilk, the Universal Extra Dimension (UED) model~\cite{Appelquist:2000nn} where one additional spatial dimension (with coordinate denoted by $y$) with a flat metric is assumed to be compactified on an $S^{1}/\mathbb{Z}_{2}$ orbifold with the compactification radius $R$ and all the SM fields accessing the bulk. The orbifold breaks the translational symmetry along the extra dimension and the end-points where this symmetry is broken are called the boundary of the orbifold. This breakdown of translational symmetry results in the violation of momentum conservation in the fifth direction, which manifests as violation of the Kaluza-Klein (KK) number. However, there remains an accidental discrete symmetry, leading to the concept of KK parity which is $(-1)^{n}$ for the $n$-th KK mode particle. This KK parity mandates the stability of the lightest KK particle, making it a suitable candidate of the dark matter (DM) and providing one strong motivation for the model.

It has been shown that radiative corrections modify both masses and couplings, partially lifting the degeneracy and paving the way for interesting collider phenomenology of the model~\cite{Cheng:2002iz}. For a recent account of collider phenomenology and relevant extensions of UED, please see Ref.~\cite{Choudhury:2016tff, Beuria:2017jez, Deutschmann:2017bth, Chakraborty:2017kjq, Arun:2018yhg, Ghosh:2018mck}.
Due to the non-renormalizability of the 5D theory, the radiative corrections involves a cutoff scale $\Lambda$. Vacuum stability studies on Higgs boson mass and couplings in the context of minimal UED (mUED) indicate $\Lambda R \sim 6$~\cite{Blennow:2011tb, Datta:2012db}. One can introduce boundary localized kinetic terms (BLKT) to accommodate such corrections~\cite{Georgi:2000ks, Carena:2002me,Flacke:2008ne, delAguila:2003bh, delAguila:2003gu, delAguila:2003gv, delAguila:2006atw}. The coefficients of these BLKTs are termed as BLKT parameters that are eventually related to the radiative corrections in a UV-complete model. In this paper, however, we will consider them as free parameters in the spirit of the so called non-minimal UED (nmUED) models. 
A multitude of phenomenological aspects of nmUED has been studied, including Large Hadron Collider (LHC) searches for the strong sector~\cite{Datta:2012tv, Datta:2013yaa}, Higgs data~\cite{Dey:2013cqa}, flavor physics~\cite{Datta:2015aka, Datta:2016flx, Biswas:2017vhc, Dasgupta:2018nzt}, DM phenomenology~\cite{Datta:2013nua, Flacke:2017xsv}, unitarity bounds~\cite{Jha:2016sre}, $Z\to b \bar b$ decay width~\cite{Jha:2014faa}, rare top decays~\cite{Dey:2016cve} and some other sectors~\cite{Datta:2012xy, Datta:2013lja, Shaw:2014gba, Shaw:2017whr}. 
Flavor-changing neutral current (FCNC) interactions are of utmost importance in constraining the BSM scenarios. In the SM, FCNCs are absent at the tree-level.  Even at the loop-level, they are strongly suppressed by the Glashow-Iliopoulos-Maiani (GIM) mechanism. This kind of loop-driven processes can get contributions from new physics particles and alter the SM predictions for these processes. Decays of the top quark induced by FCNC interactions are known to be extremely rare within the SM. However, this can be enhanced in many BSM scenarios, {\it e.g.}, 2HDM~\cite{Eilam:1990zc, Bejar:2000ub, Iltan:2001yt, Arhrib:2005nx, Abbas:2015cua, Gaitan:2017tka}, left-right symmetric model~\cite{Gaitan:2004by, Frank:2005vd}, MSSM~\cite{Guasch:1999jp, Cao:2007dk, Cao:2014udj, Dedes:2014asa}, $R$-parity violating SUSY~\cite{Eilam:2001dh, Cao:2008vk, Bardhan:2016txk}, warped extra dimensional models~\cite{Agashe:2006wa, Gao:2013fxa, Diaz-Furlong:2016ril}, UED models~\cite{GonzalezSprinberg:2007zz}, composite Higgs models~\cite{Agashe:2009di}, etc. An effective field theory based study of rare top decays can be found in~\cite{CorderoCid:2004vi, Datta:2009zb}. A number of studies were dedicated to the search of these rare decays at colliders like the LHC~\cite{DiazCruz:1989ub, Mele:1998ag, AguilarSaavedra:2002ns, AguilarSaavedra:2004wm, Chen:2013qta, Khanpour:2014xla, Hesari:2014eua, Kim:2015oua, Hesari:2015oya, Khatibi:2015aal, Malekhosseini:2018fgp, Banerjee:2018fsx}. In our previous paper~\cite{Dey:2016cve}, we considered the rare decays of $t\to c\gamma$ and $t \to ch$ in the context of nmUED. In the present paper, our aim is to explore the rare decays of $t\to cg$ and $t\to cZ$ 
in both mUED and nmUED. 
Compared to the earlier considered decay modes the main differences here are two-fold. In case of $t\to cg$, since one of the vertex contains strong coupling, the decay width is expected to be relatively larger compared to the other rare decay modes that are considered. The intent is to explore if that holds in the (n)mUED case. In case of $t\to cZ$, apart from the fact that one of the final state particle is one massive gauge boson $Z$ instead of previously studied massless $\gamma$, there will be a few more Feynman diagrams compared to $t\to c\gamma$ owing to the fact that $Z$ will have additional couplings with KK charged Higgses and $W$s, which the $\gamma$ does not have.

The paper is organized as follows. In Section~\ref{sec:model}, we briefly describe the minimal and non-minimal version of the UED model. We just highlight the main ingredients, leaving finer details to Ref.~\cite{Dey:2016cve}. Section~\ref{sec:raredecays} is dedicated to discussions about general features of the $t\to cg$ and $t\to cZ$ decays in these models. We elaborate on the results in the SM, (n)mUED in Section~\ref{sec:results}. In Section~\ref{sec:constrnts}, we discuss the compatibility of parameter space with phenomenological constraints coming from electroweak precision data, LHC observations, and observables associated with the down sector. Finally, we summarize and conclude in Section~\ref{sec:concl}. Feynman rules relevant to this study are listed in the appendix.

\section{Model}
\label{sec:model}

In this section, we will follow the same notational convention as in Ref.~\cite{Dey:2016cve}. We therefore refer to that paper as well as~\cite{Flacke:2008ne, Datta:2013nua, Dvali:2001gm, Carena:2002me, delAguila:2003bh, delAguila:2003gu, delAguila:2003gv, delAguila:2006atw, Datta:2012tv, Datta:2013yaa, Datta:2012xy, Datta:2013lja, Dey:2013cqa, Jha:2014faa, Shaw:2014gba, Datta:2015aka, Jha:2016sre, delAguila:2003kd, Flacke:2014jwa} for the essential features of the model. In this section, we very briefly review the model and point out a few details that were skipped in~\cite{Dey:2016cve}.

\subsection{Lagrangian and Interactions}
\label{s:Lag_nmUED}

We consider the extra spatial dimension ($y$) to be compactified on an $S^{1}/\mathbb{Z}_{2}$ orbifold with the fixed points located at $y = 0$ and $y = \pi R$, where $R$ denotes the compactification radius. The five-dimensional action for a generic fermion field $\mathscr{F}$ can be written as
\begin{eqnarray}
\label{actn_quark}
\mathcal{S}_{{\rm fermion}} = \int d^4 x \int_{0}^{\pi R} dy \Big[\overline{\mathscr{F}} i\Gamma^{M} \mathcal{D}_{M} \mathscr{F} + r_f \{ \delta(y) + \delta(y-\pi R) \} \overline{\mathscr{F}} i\gamma^{\mu} \mathcal{D}_{\mu} P_{L/R} \mathscr{F}\Big], 
\end{eqnarray}
where the Latin indices run from 0 to 4 and the Greek indices from 0 to 3. The five-dimensional Gamma matrices are $\Gamma^{M}=(\gamma^{\mu},-i\gamma_{5})$. The covariant derivative is given by 
\begin{equation}
\mathcal{D}_M\equiv \partial_M 
             - i \widetilde{g_{s}}G_M^a T^a 
             - i \widetilde{g}W_M^a \tau^a 
             - i \widetilde{g}^\prime B_M Y,
\end{equation}
where $\widetilde{g_{s}}$, $\widetilde{g}$ and $\widetilde{g}^\prime$ are the 5-dimensional gauge coupling constants of the $SU(3)_C$, $SU(2)_L$ and $U(1)_Y$ groups, respectively, and $T^a$, $\tau^a$ and $Y$ are the corresponding generators.

Here we consider $\mathscr{F}(x,y) \in \{Q_{t,b}(x,y), U(x,y), D(x,y)\}$, and the four-component five-dimensional fields are comprised of chiral spinors and their Kaluza-Klein excitations. They can be written as
\begin{subequations}
\begin{align}
Q_{t,b}(x,y) &=  N_{Q0}\,Q_{t,b L}^{(0)}+ \sum^{\infty}_{n=1}\left\lbrace Q_{t,b L}^{(n)}(x) f_{L}^{(n)}(y) + Q_{t,b R}^{(n)}(x) g_{L}^{(n)}(y) \right\rbrace, \\
U(x,y) &= N_{Q0}\,U_{R}^{(0)}+ \sum^{\infty}_{n=1}\left\lbrace U_{L}^{(n)}(x) f_{R}^{(n)}(y) + U_{R}^{(n)}(x) g_{R}^{(n)}(y)\right\rbrace, \\
D(x,y) &= N_{Q0}\,D_{R}^{(0)}+ \sum^{\infty}_{n=1}\left\lbrace D_{L}^{(n)}(x) f_{R}^{(n)}(y) + D_{R}^{(n)}(x) g_{R}^{(n)}(y)\right\rbrace.
\end{align} 
\end{subequations}
In the effective four-dimensional theory, the zero modes of $Q$ will give rise to the $SU(2)_{L}$ doublet quarks whereas the zero mode of $U$ ($D$) will be identified with the up- (down-) type singlet quarks. Using the third generation quarks as an example, after compactification and orbifolding the zero modes of $Q$ are the left-handed doublet comprising $t_{L}$ and $b_{L}$, whereas $t_{R}$ and $b_{R}$ would emerge from $U$ and $D$, respectively. Here $N_{Q0}$ is the normalization constant for fermionic wave functions of the zero modes. In Eq.~\eqref{actn_quark}, the terms with the Dirac $\delta$-functions are called the boundary localized kinetic terms (BLKTs) as they are localized in the orbifold fixed points $y = 0, \pi R$ and the corresponding parameter $r_{f}$ is the fermion BLKT parameter. By setting BLKT parameters in the nmUED model to zero, one can recover the mUED model. Note that we will take a universal fermionic BLKT parameter in this paper.

The explicit forms of the mode functions $f_{L,R}(y), ~ g_{L,R}(y)$, normalization factors, orthonormalization of the mode functions and KK mass-determining transcendental equations are given in Section 2.1 of Ref.~\cite{Dey:2016cve}. The five-dimensional actions for gauge and scalar fields, Yukawa interactions can be written in the same spirit as the fermions with boundary localized parameters, $r_{g}, r_{\Phi}$, and $r_{y}$ respectively. We again refer to~\cite{Dey:2016cve} for further details.

We now discuss gauge fixing that is not elaborated in~\cite{Dey:2016cve}. We use the 't Hooft--Feynman gauge in our calculations, and the gauge fixing action for the $W$ boson for example is given by~\cite{Muck:2004zz, Jha:2014faa, Datta:2014sha},
\begin{align}
\label{gauge_fix}
%
%
\mathcal{S}_{\rm GF}^{W} = -\frac{1}{\xi _y}\int d^{4}x\int_{0}^{\pi R} dy \Big\vert\partial_{\mu}W^{\mu +} + \xi_{y}\left(\partial_{5}W^{5+} 
-iM_{W}\phi^{+}\{1 + r_{\Phi}\left(\delta(y) + \delta(y - \pi R)\right)\}\right)\Big \vert^{2},
\end{align}
where $M_W$ is the mass of the $W$ boson. The $y$-dependent gauge fixing parameter $\xi_y$ is related to the gauge fixing parameter $\xi$ (equals to $1$ in Feynman gauge, and $0$ in Landau gauge) by \cite{Jha:2014faa, Datta:2015aka, Dey:2016cve, Muck:2004zz}
\begin{equation}
\frac{1}{\xi_{y}}= \frac{1}{\xi}\{1 + r_{\Phi}\left(\delta(y) + \delta(y - \pi R)\right)\}.
\end{equation}
The above-mentioned gauge-fixing action prohibits the mixing between $W_{\mu}^{(n)\pm}$ and $W_{5}^{(n)\pm}$ and also between $W_{\mu}^{(n)\pm}$ and $\phi^{(n)\pm}$. 

We also note in passing a few key aspects of this model. The standard procedure to obtain the effective four-dimensional couplings is to write the original five-dimensional interaction terms with each field replaced by their corresponding KK expansions and then integrate out the extra dimension $y$. In mUED, this type of couplings are similar to their SM counterparts. But in the case of nmUED, the couplings get further modifications from the overlap integrals which have the following generic form:
\begin{align}
I^{ijk} = \int_{0}^{\pi R}dy~f_{\alpha}^{(i)}(y)
              ~f_{\beta}^{(j)}(y)~f_{\gamma}^{(k)}(y),
\label{eq:oi}
\end{align}      
where the Latin indices (superscripts) refer to the KK-level of the respective fields and the Greek indices (subscripts) denote the type of fields. This modification in coupling is a novel feature of the nmUED scenario. The reason of this modification is that unlike mUED, the KK-mode function in nmUED has BLKT parameter dependence, explicitly in the normalization factors and implicitly in the KK masses. Also note that the conservation of KK parity\footnote{Note that in nmUED as long as the BLKT parameters are the same at the two boundaries $y = 0$ and $y = \pi R$ , KK parity is conserved. Also, this is easy to see. KK parity is actually a translational symmetry under the transformation, $y \to y-\pi R$. Since for symmetric BLKTs, i.e., the BLKT parameters are the same at the two boundaries, we have 
$
r\{\delta(y)+\delta(y-\pi R)\} \to r\{\delta(y-\pi R)+\delta(y-2\pi R)\} 
 =r\{\delta(y-\pi R)+\delta(y)\}.
$
Thus KK parity remains invariant. Had there been asymmetric BLKTs, KK parity would have been broken,
$
\{r_{1}\delta(y)+r_{2}\delta(y-\pi R)\} \to \{r_{1}\delta(y-\pi R)+r_{2}\delta(y-2\pi R)\} 
 =\{r_{1}\delta(y-\pi R)+r_{2}\delta(y)\}.
$} ensures that these overlap integrals vanish if $(i+j+k)$ is an odd integer.

In our analysis, we make a few simplified assumptions in our choice of the BLKT parameters. We take a universal parameter $r_{\Phi}$ for all the bosonic fields and a universal $r_{f}$ for all the fermionic fields. Also, we take the boundary localized Yukawa parameter $r_{y}$ to be equal to $r_{f}$. Although these choices can be generalized and a number of new parameters can be introduced, this minimal setup comes with a few advantages, such as the absence of FCNC's owing to the flavor-blind fermion BLKT parameter, a simpler fermion mixing matrix (to be shown later), and a flat zero mode of the Higgs field~\cite{Flacke:2013pla, Datta:2013yaa}. Moreover, in our numerical analysis we avoid negative BLKT parameters which can give rise to tachyonic and ghost fields in the theory.

\subsection{Physical Eigenstates}
\label{sbsc:Physcl_Eigenst}

In the effective four-dimensional theory, the higher KK modes of various fields will mix among one another to give the physical fields. This is generic to the fermionic as well as the scalar/gauge sectors. In Section 2.2 of Ref.~\cite{Dey:2016cve} we discussed the details of the physical eigenstates of fermions, gauge bosons, and scalars. We just review the key points here. 
%

%
%
For the choice of $r_f = r_y$ used in this analysis, one gets a simpler form for the fermion mixing matrix and avoids the mode mixing. The gauge eigenstates $Q_{j}^{(n)}$ ($D^{(n)}$) and mass eigenstates $Q_{j}^{\prime (n)}$ ($D^{\prime (n)}$) (in this notation, $j$ refers to the down-type quark flavor) are related as follows:
\begin{subequations}
\begin{align}
Q_{j_{L/R}}^{(n)} &= \mp \cos \alpha_{n} Q_{j_{L/R}}^{\prime(n)} +
                      \sin \alpha_{n} D_{L/R}^{\prime(n)},\\
D_{L/R}^{(n)} &= \pm \sin \alpha_{n} Q_{j_{L/R}}^{\prime(n)} + 
                    \cos \alpha_{n} D_{L/R}^{\prime(n)},
\end{align}
\end{subequations}
where the mixing angle $\alpha_{n} = \frac12 \tan^{-1}(m_{b}/M_{Qn})$.
The mass eigenstates share the same mass eigenvalue,  
\begin{align}
m_{Q_{b}^{\prime (n)}} = m_{D^{\prime (n)}} 
                       = \sqrt{m_{b}^{2}+M_{Qn}^{2}} 
                          \equiv M_{\rm bottom}.
\end{align}
An analogous result applies to the up sector.

The gauge and scalar sectors are intertwined, in the sense that the fifth components of the gauge fields mix with the appropriate scalar KK degrees of freedom and form their physical scalars. 
The mixing between the fifth components of $W^{\pm}$ and the KK excitations of $\phi^{(0)\pm}$ of the Higgs doublet field render charged KK Goldstone modes, $G^{\pm (n)}$ and charged KK scalars $H^{\pm (n)}$ ~\cite{Petriello:2002uu}. Explicitly, 
\begin{subequations}
\label{eq:Gn_Hn}
\begin{align}
G^{\pm (n)} &= \frac{1}{M_{W_{n}}}\left(M_{\Phi n}W^{\pm5(n)}\mp  iM_{W}\phi^{\pm(n)}\right),\\
H^{\pm (n)} &= \frac{1}{M_{W_{n}}}\left(M_{\Phi n}\phi^{\pm(n)}\mp iM_{W}W^{\pm5(n)}\right).
\end{align}
\end{subequations}
The fields $W^{\mu (n)\pm}$, $G^{(n)\pm}$ and $H^{(n)\pm}$ share the same mass eigenvalue $M_{W n} \equiv \sqrt {M_{\Phi n} ^2 + M_W^2}$ in the 't-Hooft Feynman gauge.
The above combinations of charged Higgs and charged Goldstone modes ensure a vanishing coupling for interactions of the form $\gamma^{\mu (0)} H^{(n) \pm}W_{\nu}^{(n) \mp}$. Surely, different convention in the definition of the field strength will alter the combinations in Eqs.~\eqref{eq:Gn_Hn} that will make this coupling vanishing\footnote{An important comment about the sign convention is in order. The combinations are dependent on convention of the sign used before the non-Abelian part of the field strength tensor $\mathcal{F}_{MN}^{a}$; the couplings required for the above combination comes from $\left(\mathcal{D}^{\mu}\Phi\right)^{\dagger}\left(\mathcal{D}_{\mu}\Phi\right)$ and $\mathcal{F}_{\mu 5}^{a}\mathcal{F}^{\mu 5 a}$. The expressions of the charged Higgs and charged Goldstone modes do not depend on the sign used in $\left(\mathcal{D}^{\mu}\Phi\right)^{\dagger}\left(\mathcal{D}_{\mu}\Phi\right)$, but depend on whether $\mathcal{F}_{MN}^{a}$ is $\left(\partial_{M}\mathcal{W}_{N}^{a}-\partial_{N}\mathcal{W}_{M}^{a}+\tilde{g}f^{abc}\mathcal{W}_{M}^{b}\mathcal{W}_{N}^{c}\right)$ or $\left(\partial_{M}\mathcal{W}_{N}^{a}-\partial_{N}\mathcal{W}_{M}^{a}-\tilde{g}f^{abc}\mathcal{W}_{M}^{b}\mathcal{W}_{N}^{c}\right)$. For  $\mathcal{F}_{MN}^{a}\equiv\left(\partial_{M}\mathcal{W}_{N}^{a}-\partial_{N}\mathcal{W}_{M}^{a}-\tilde{g}f^{abc}\mathcal{W}_{M}^{b}\mathcal{W}_{N}^{c}\right)$, the combinations which give the vanishing coupling of  $A^{\mu (0)} H^{(n) \pm}W_{\nu}^{(n) \mp}$ are $$ G^{\pm (n)} = \frac{1}{M_{W_{n}}}\left(M_{\Phi n}W_{5}^{\pm(n)}\mp  iM_{W}\phi^{\pm(n)}\right), $$$$H^{\pm (n)} = \frac{1}{M_{W_{n}}}\left(M_{\Phi n}\phi^{\pm(n)}\mp iM_{W}W_{5}^{\pm(n)}\right).$$}.
%
%

\section{Rare Decays}
\label{sec:raredecays}

In the SM, the rare $t\to cg$ and $t\to cZ$ decays are highly suppressed due to the CKM factor, the GIM mechanism as well as the loop factor, making the branching ratios of these channels minuscule. In the BSM models, however, new physics particles can contribute in these loop-driven processes and may increase the branching ratios orders of magnitude higher than the SM expectations. In this section, we discuss these decays in the model. We start by discussing the general Lorentz structure and Feynman diagrams of each of these decays. The relevant Feynman rules in the 't Hooft--Feynman gauge are presented in the Appendix of~\cite{Dey:2016cve}. We have made use of \texttt{Package-X}~\cite{Patel:2015tea} and \texttt{LoopTools}~\cite{Hahn:1998yk} in the numerical computation.

\subsection{$t\to cg$}
\label{sbsc:tcg}

\begin{figure}[t]
  \centering
  \subfloat[]{
    \includegraphics[scale=0.35]{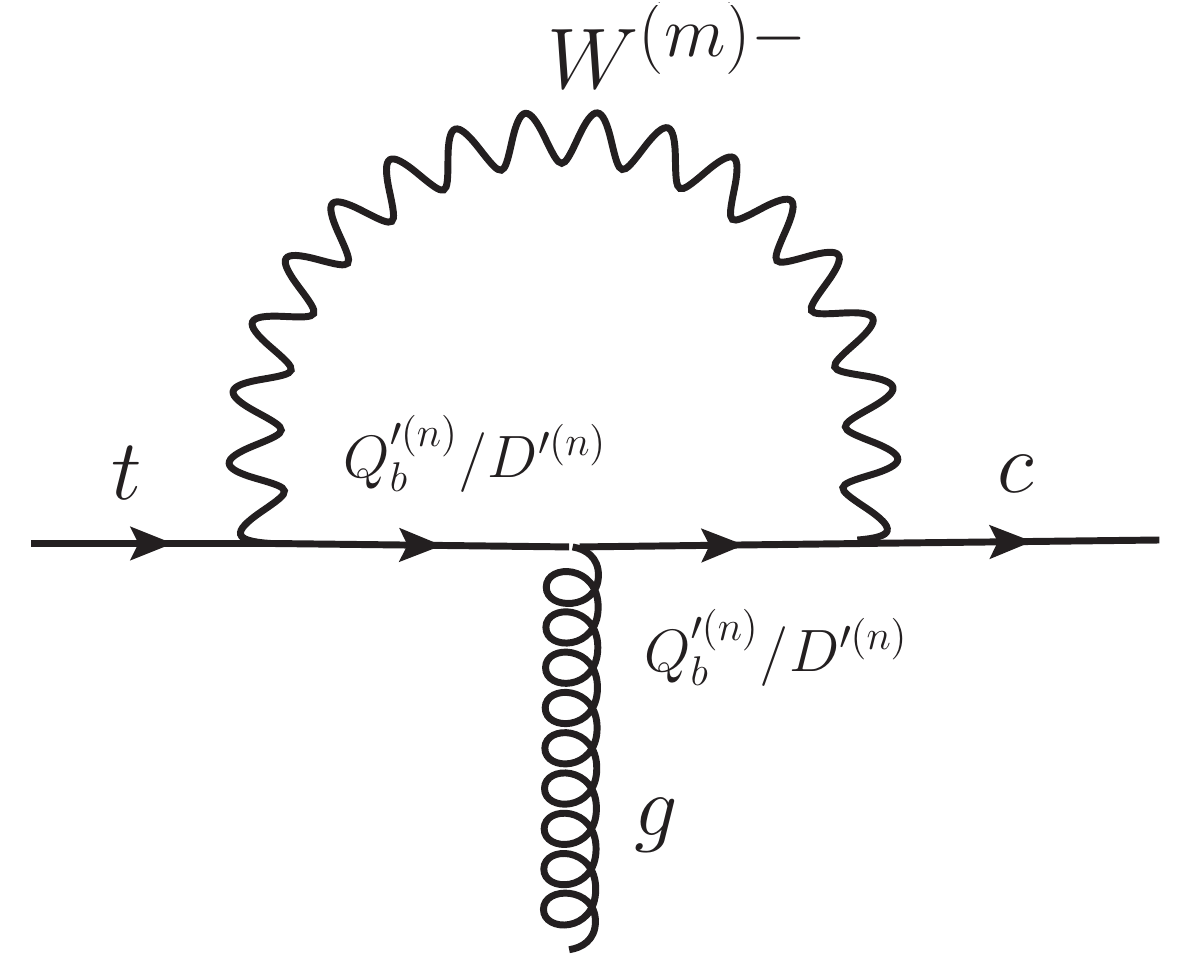}
    }
  \subfloat[]{
    \includegraphics[scale=0.35]{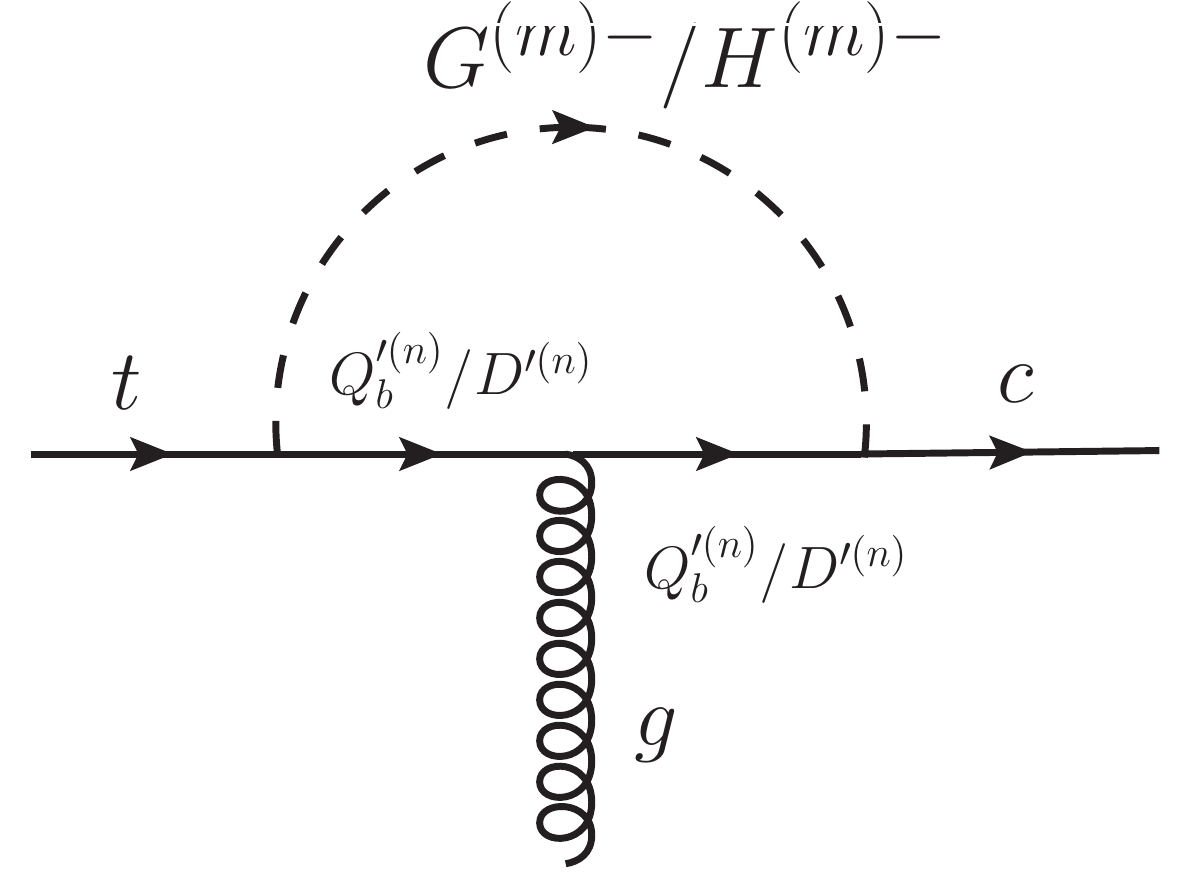}
    }
  \subfloat[]{
    \includegraphics[scale=0.35]{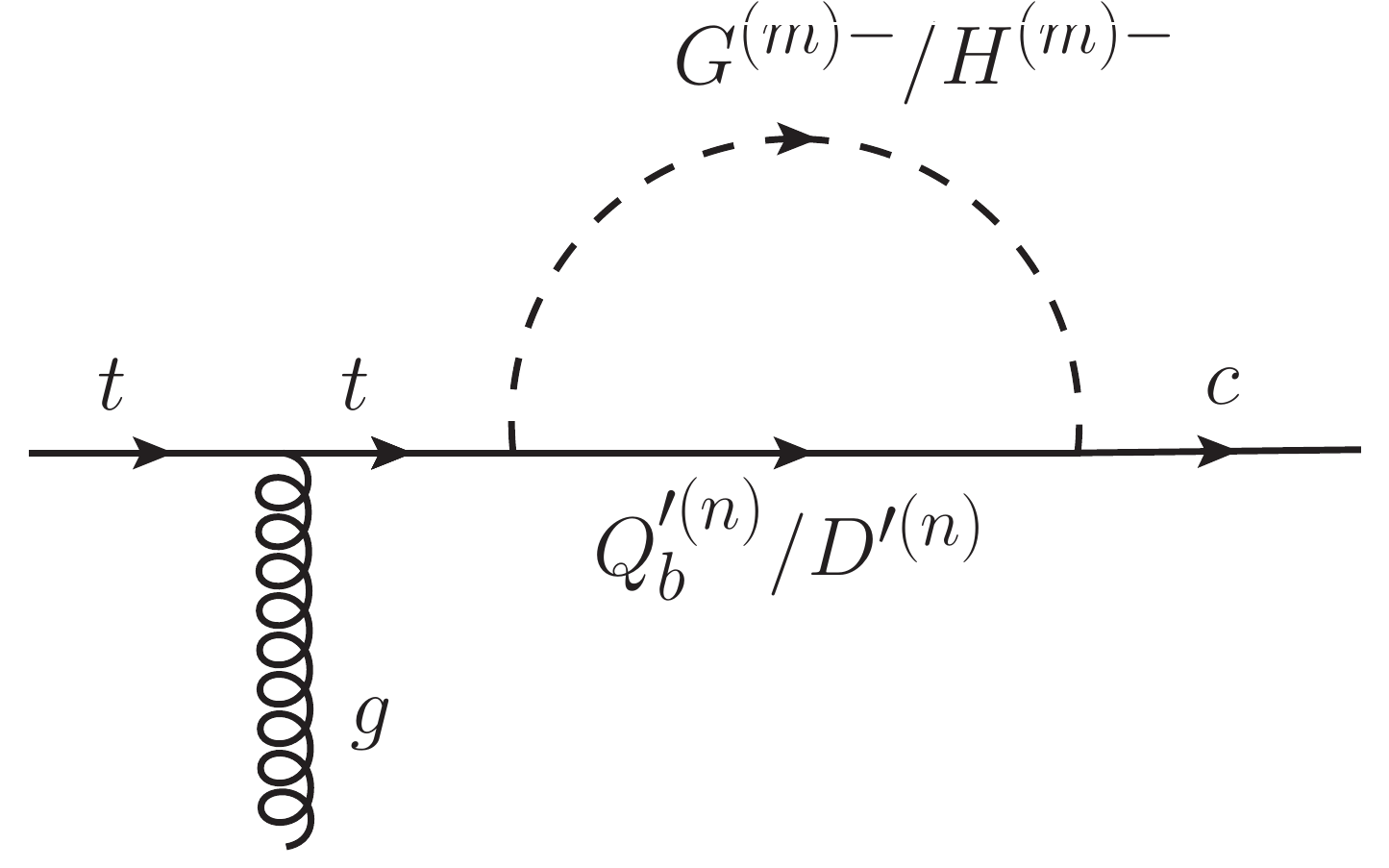}
    } \\
  \subfloat[]{
    \includegraphics[scale=0.35]{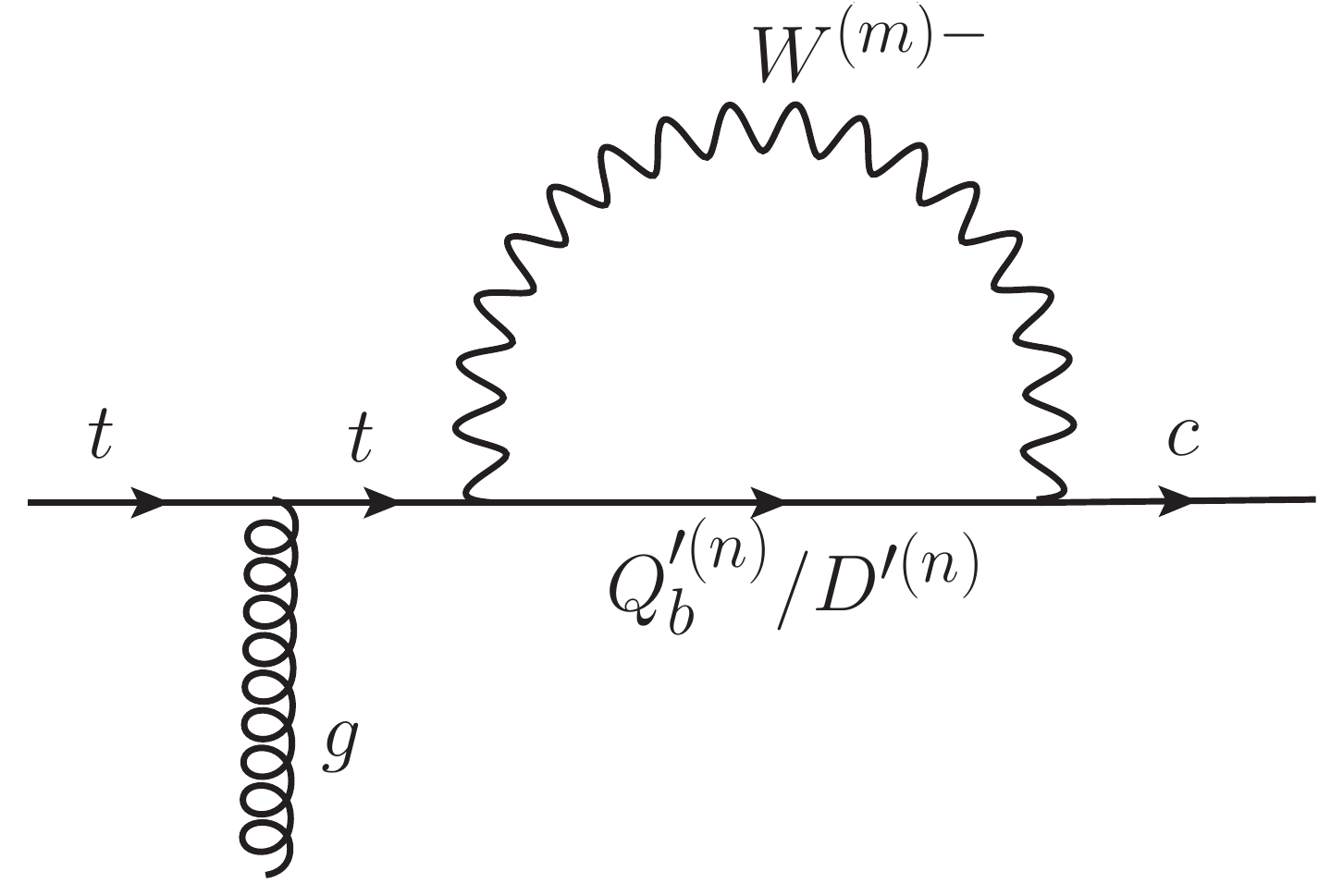}
    }
  \subfloat[]{
    \includegraphics[scale=0.35]{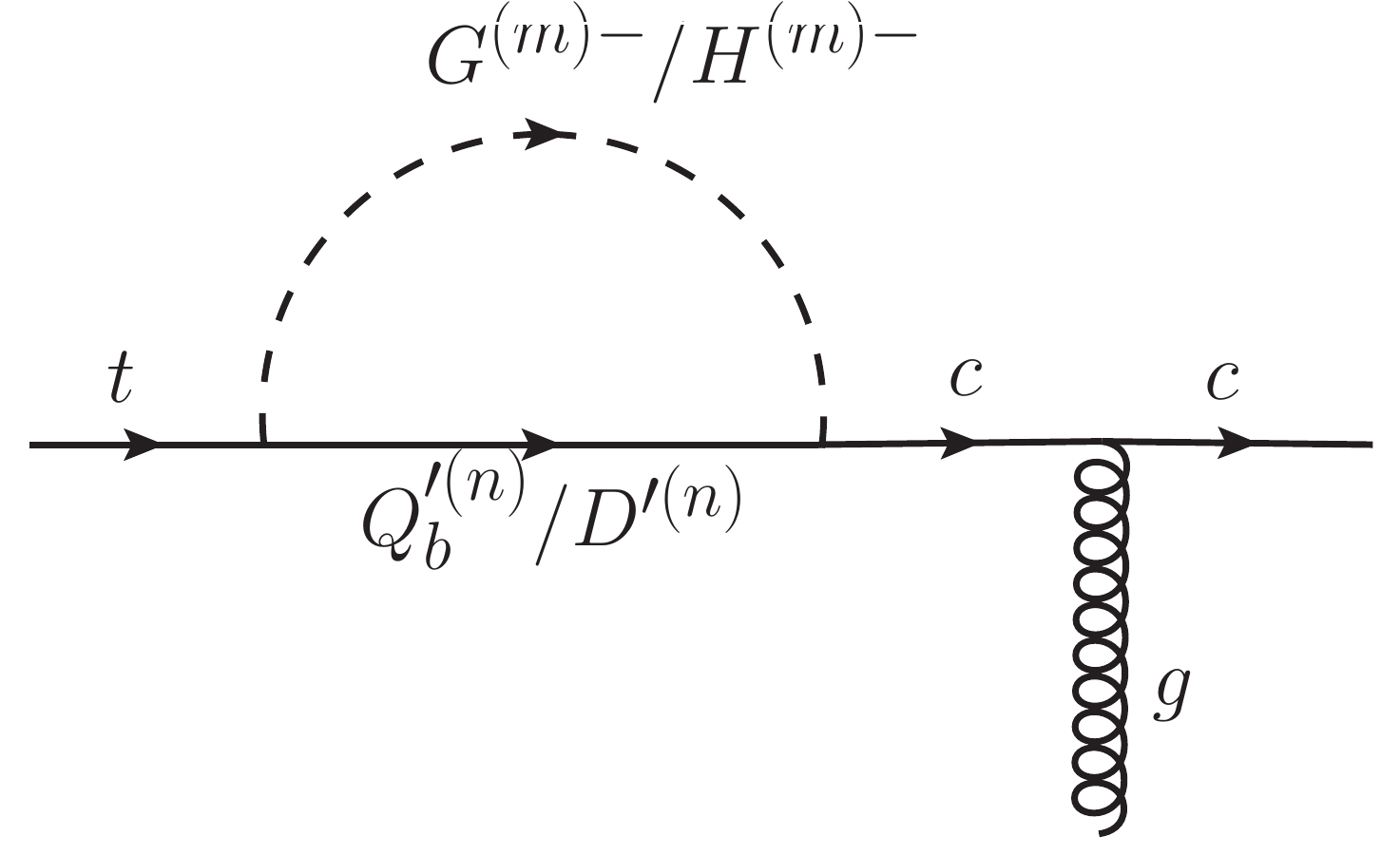}
    } \\
  \subfloat[]{
    \includegraphics[scale=0.35]{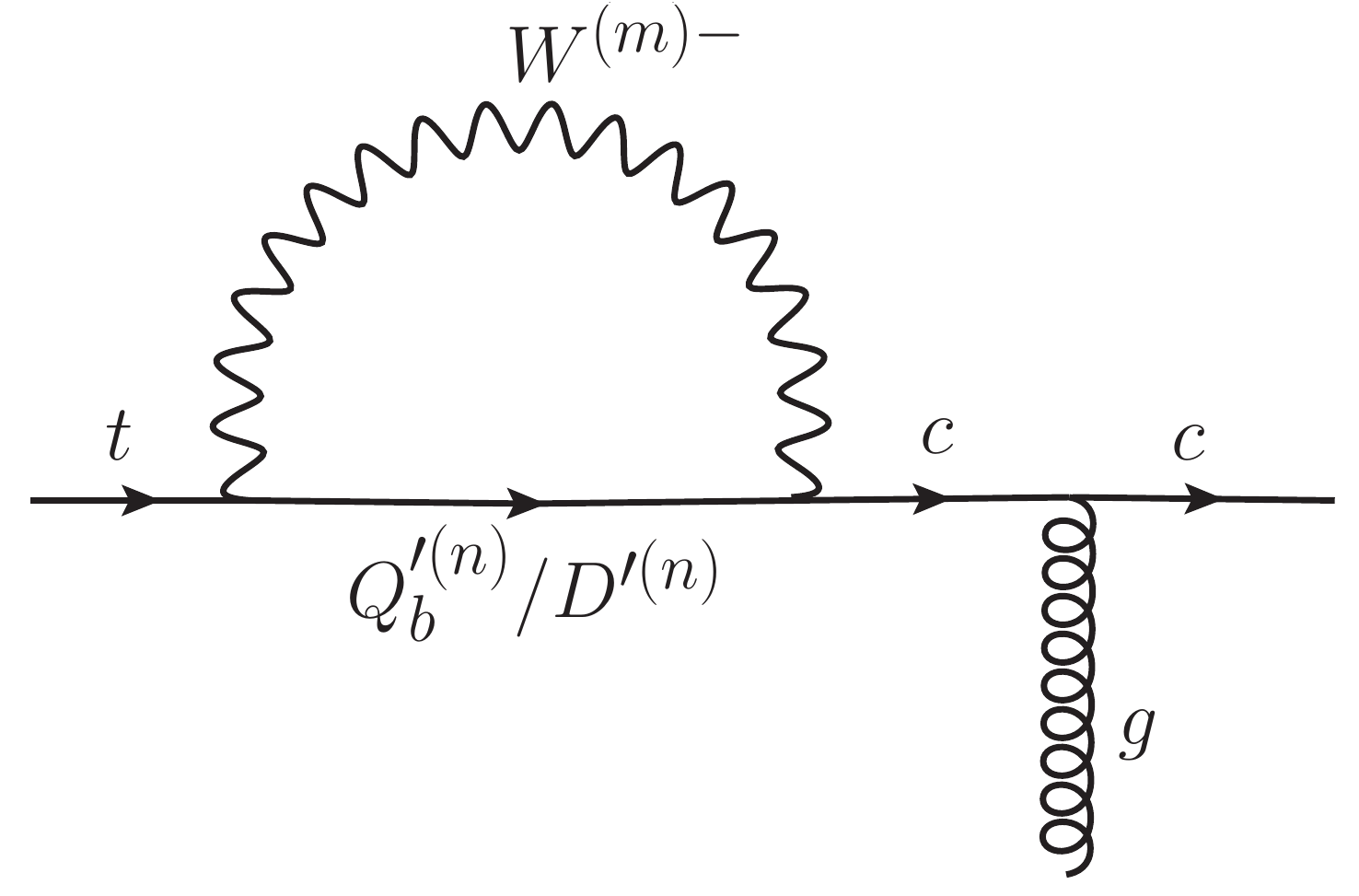}
    } 
    \caption[]{Feynman diagrams for the process $t\to cg$ in the 't Hooft--Feynman gauge in the nmUED model. Note that index $n$ can also be zero provided they are allowed by the KK parity.  Such diagrams are also included in our numerical calculations.
    }
\label{fig:tcg}
\end{figure}

The most general form of the amplitude of the $t(p)\to c(k_{2})g(k_{1})$ decay for on-shell quark and real gluon can be written as~\cite{AguilarSaavedra:2002ns}
\begin{align}
\mathcal{M}(t\to cg) = \frac{i}{m_{t}+m_{c}}
                       \bar{u}(k_{2})
                       [\sigma^{\mu \nu} k_{1\nu}
                       \left(A_{L}P_{L}+B_{R}P_{R}\right)]
                       u(p)
                       \epsilon^{\ast}_{\mu}(k_{1}).
\end{align}
The information of couplings, CKM matrix elements and the loop integrals are encoded in the coefficients $A_{L}$ and $B_{L}$. 
Many of the features of $\mathcal{M}(t\to cg)$ are identical to those of $\mathcal{M}(t\to c \gamma)$. We do not repeat such features here as they have been expatiated in Sec. 3.1 of~\cite{Dey:2016cve}. The decay width in general is given by
\begin{align}
\Gamma_{t\to cg} = \frac{C_{F}}{16\pi}
                   \frac{(m_{t}^{2}-m_{c}^{2})^{3}}
                   {m_{t}^{3}(m_{c}+m_{t})^{2}}
                   \left(|A_{L}|^{2} + |B_{R}|^{2}\right),
\end{align}
where $C_{F} = 4/3$ is a colour factor. The relevant Feynman diagrams are shown in Fig.~\ref{fig:tcg}, where the superscript $(n)$ or $(m)$ represents the KK number. In mUED, the KK number is conserved at any specific vertex and, therefore, we always have $m = n$.  In the nmUED case, nevertheless, $m$ and $n$ can be different respecting the KK parity is conserved. The quantities $A_L$ and $B_R$ contain the summation over the KK modes running in the loops. In our numerical calculations, we take the sum up to the fifth KK level as the contribution of higher modes decouples. (We have checked that the results are virtually unchanged when we sum up to the tenth KK level.) Also, $m_c$ plays an insignificant role in the total decay width. Unless otherwise stated, we neglect $m_c$ in our numerical analysis.

\subsection{$t\to cZ$}
\label{sbsc:tcZ}

We now consider the $t(p) \to c(k_{2}) Z(k_{1})$ decay, the amplitude of which in its most general form is given by~\cite{Datta:2009zb}
\begin{align}
\mathcal{M}(t\to cZ) = \bar{u}(k_{2})
                       \left[A_{L}\gamma^{\mu}P_{L} 
                           + B_{R}\gamma^{\mu}P_{R} 
                           + \frac{i \sigma^{\mu \nu}k_{1\nu}}
                                  {m_{t}+m_{c}}
                             \left(\tilde{A}_{L}P_{L} 
                                +  \tilde{B}_{R}P_{R}\right)
                       \right] 
                       u(p) \epsilon^{\ast}_{\mu}(k_{1}).                          
\end{align}
As in the previous case, here the information of couplings, CKM matrix elements, and loop momenta integrals are embedded in the coefficients $A_{L},~B_{R},~\tilde{A}_{L}$, and $\tilde{B}_{R}$. In case of $m_c=0$, $B_{R}, \tilde{A}_{L}=0$ identically. The general feature of divergence cancellation remains the same as that of $t\to cg$.
\begin{figure}[!htbp]
  \centering
  \subfloat[]{
    \includegraphics[scale=0.3]{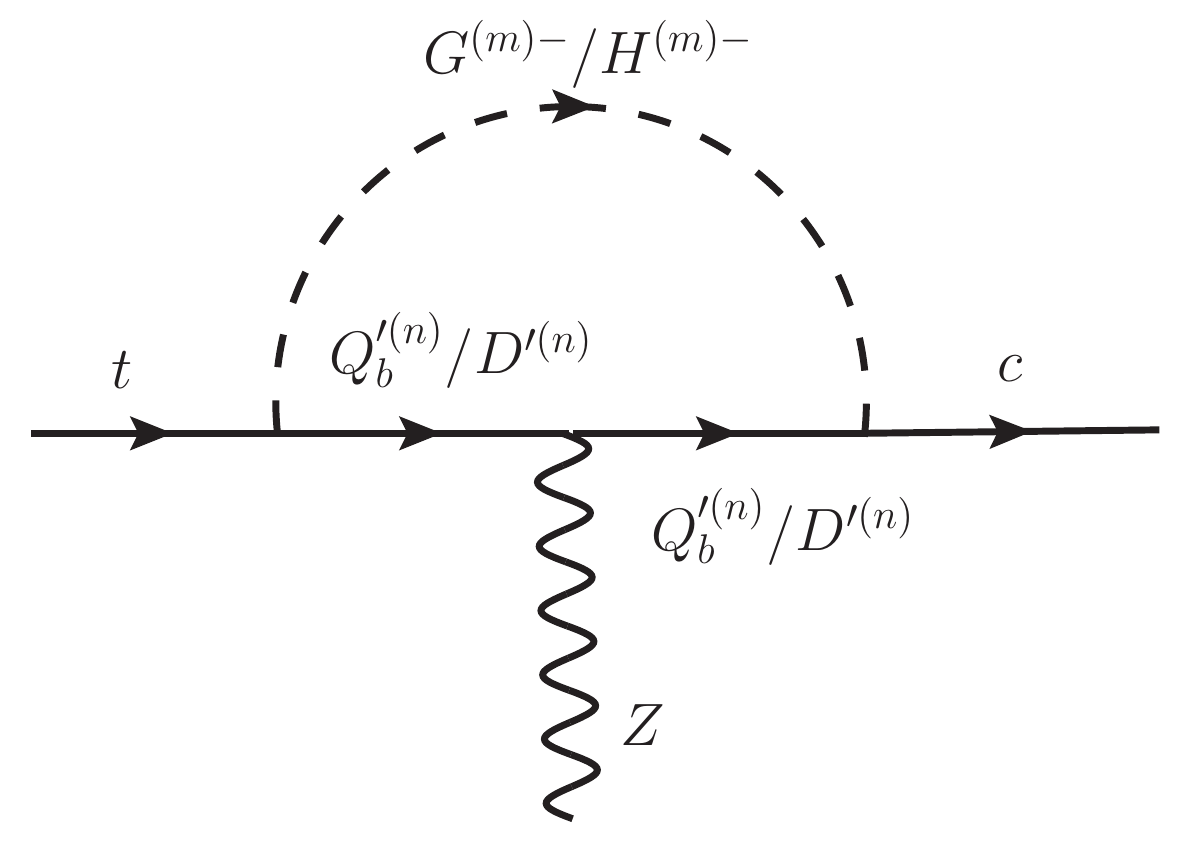}
    }
  \subfloat[]{
    \includegraphics[scale=0.3]{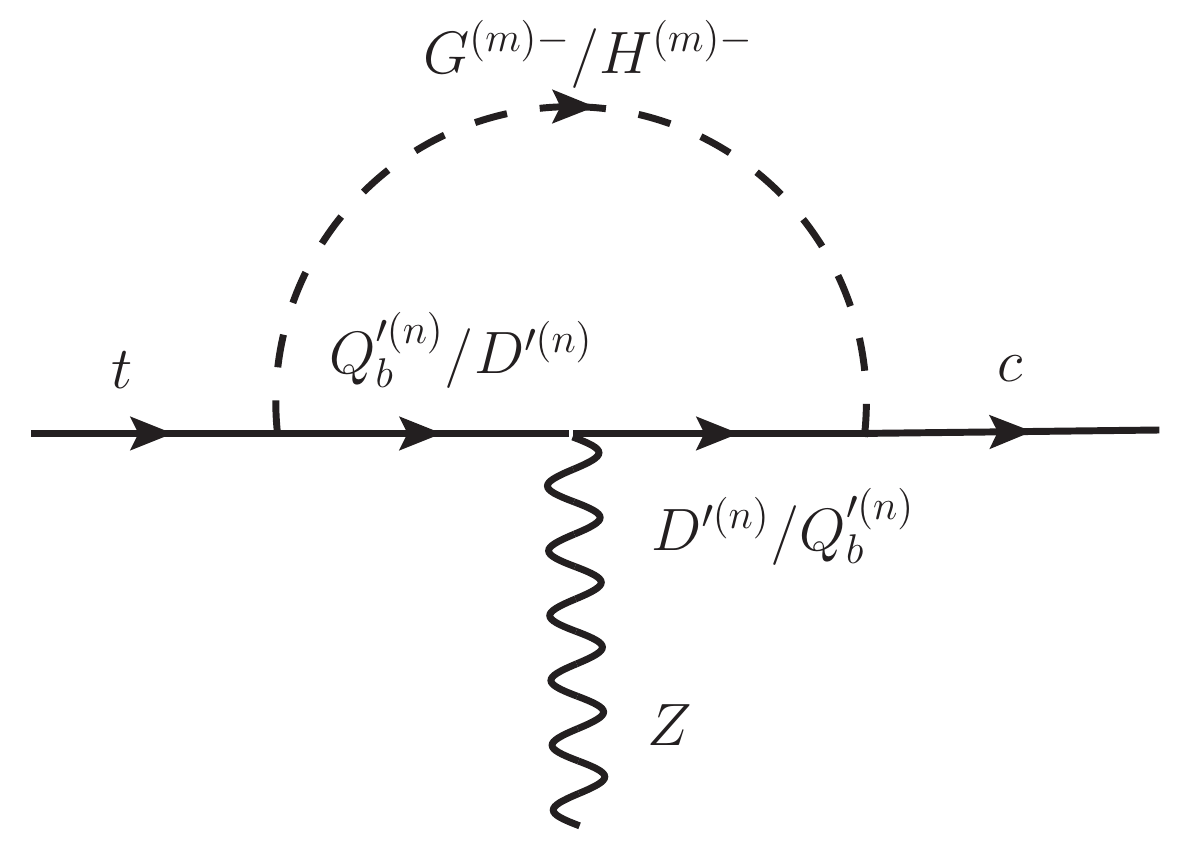}
    }
  \subfloat[]{
    \includegraphics[scale=0.3]{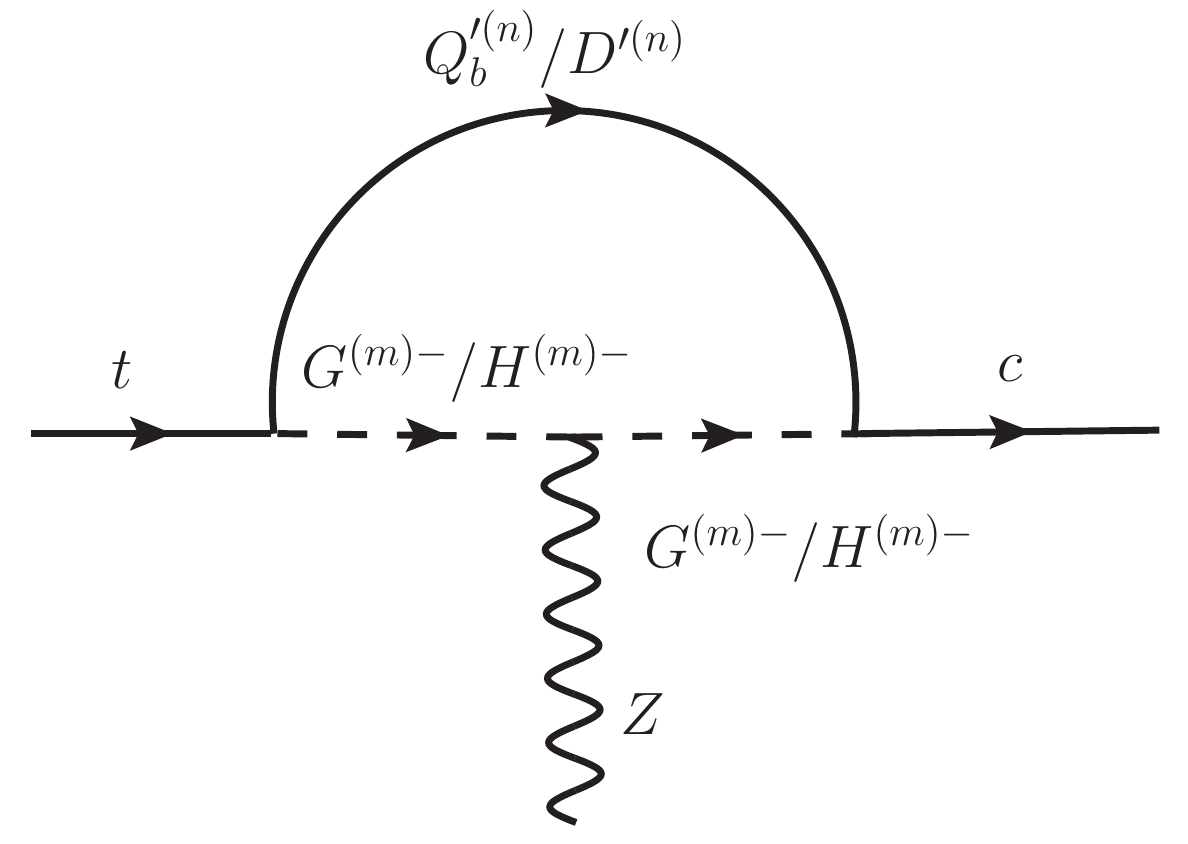}
    } \\
  \subfloat[]{
    \includegraphics[scale=0.3]{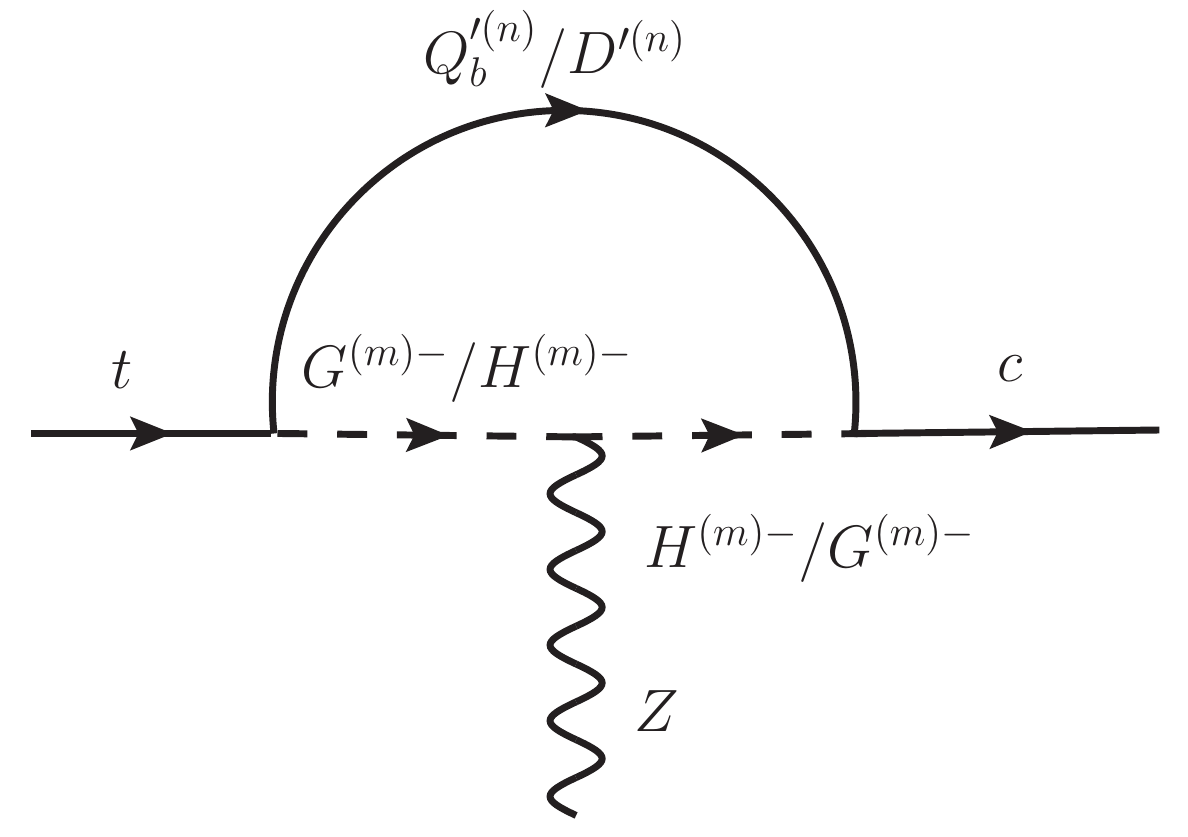}
    }    
  \subfloat[]{
    \includegraphics[scale=0.3]{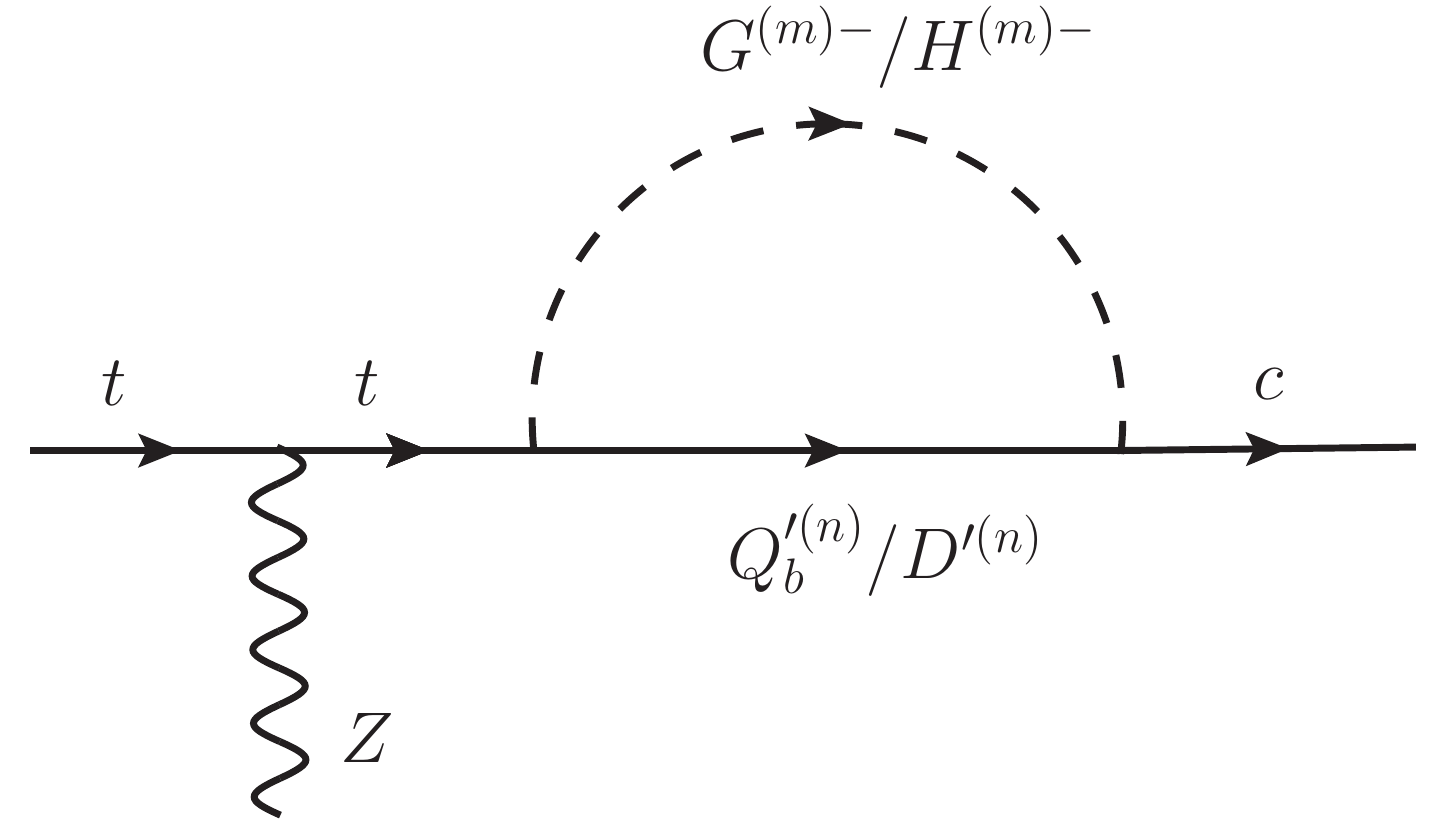}
    } 
  \subfloat[]{
    \includegraphics[scale=0.3]{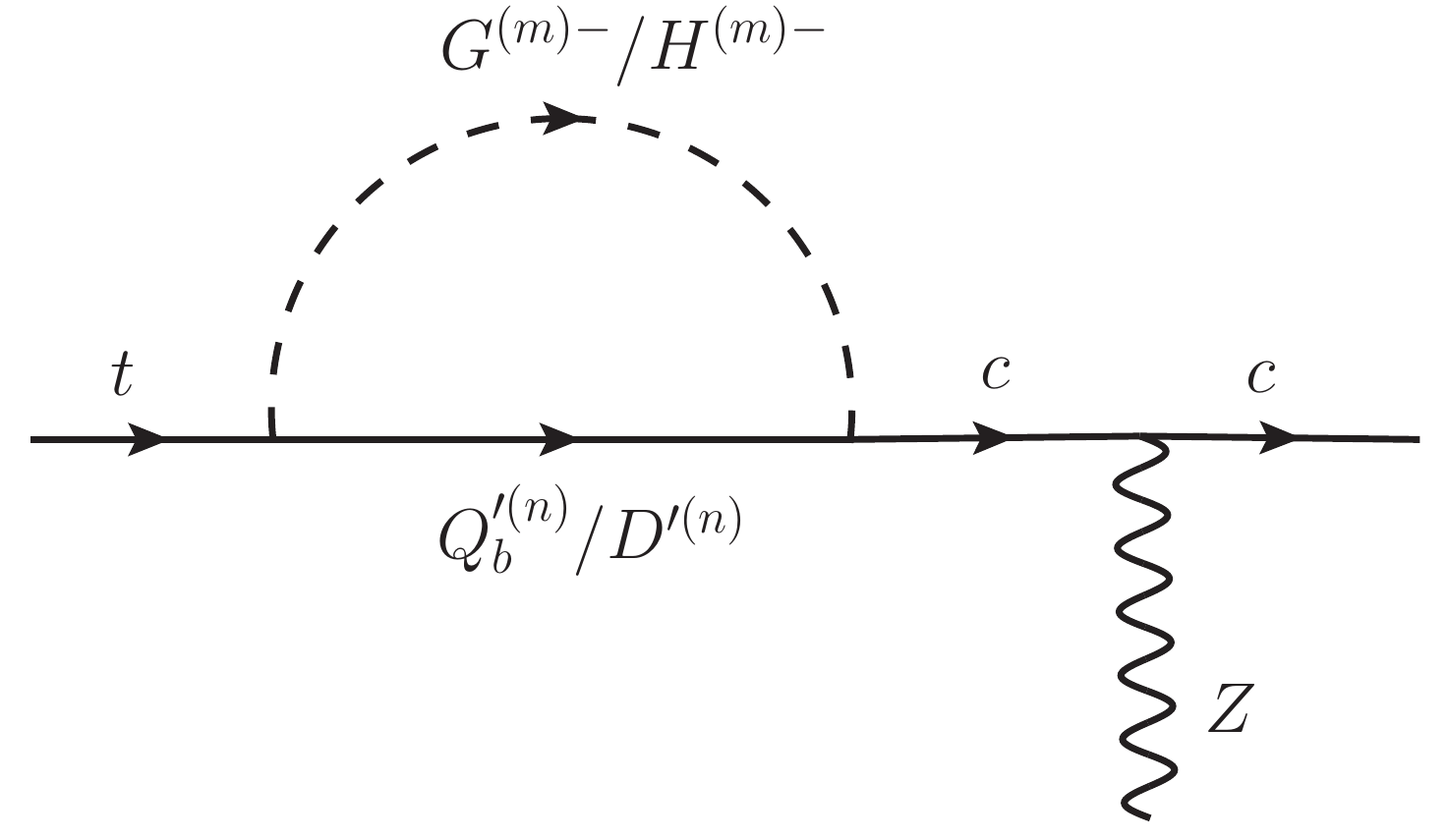}
    } \\
  \subfloat[]{
    \includegraphics[scale=0.3]{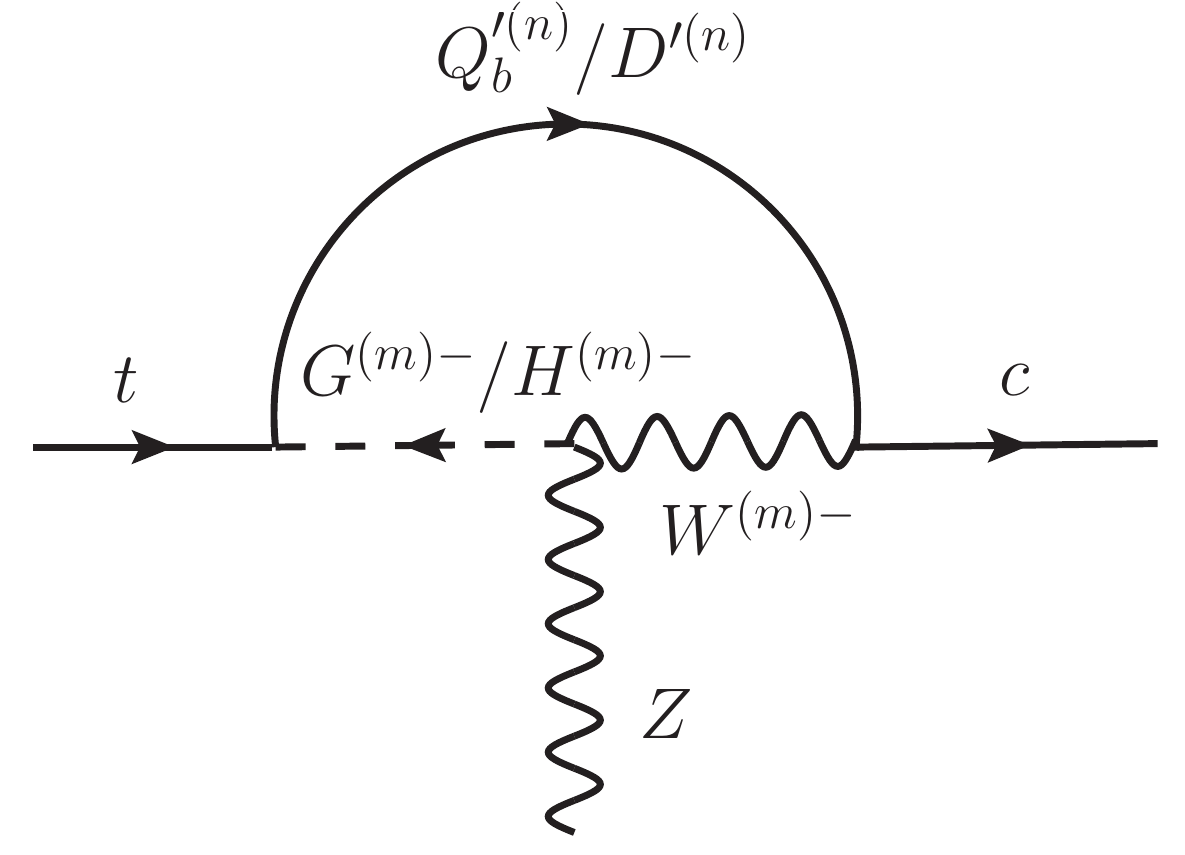}
    }
  \subfloat[]{
    \includegraphics[scale=0.3]{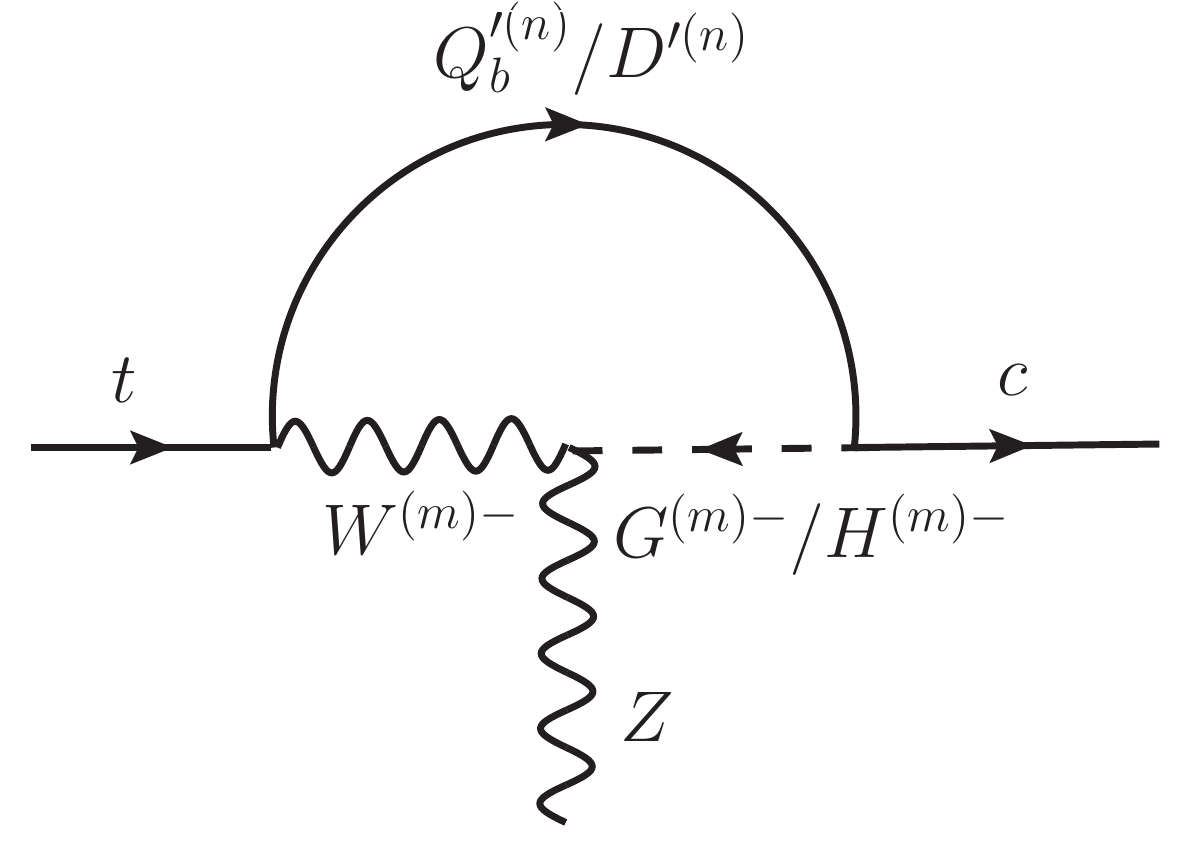}
    } 
  \subfloat[]{
    \includegraphics[scale=0.3]{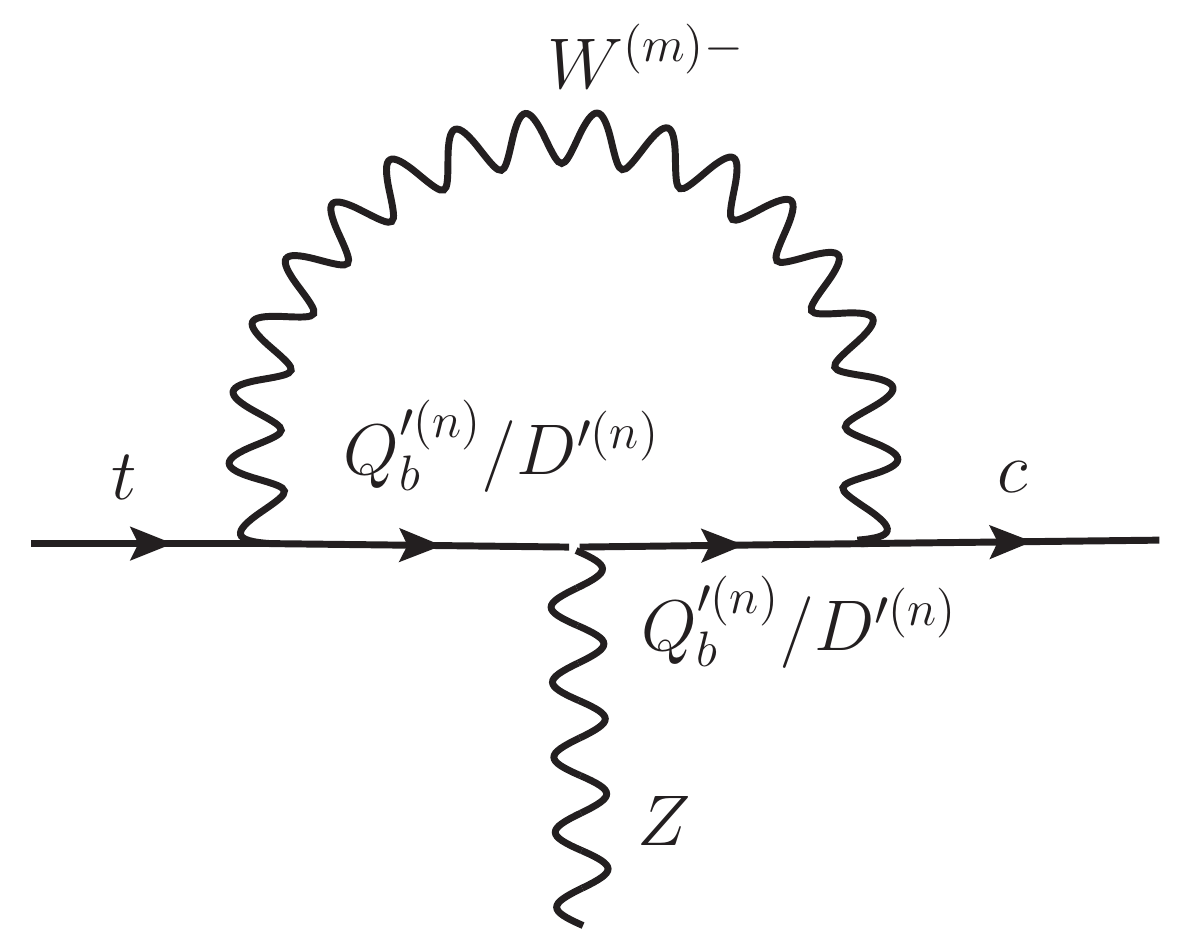}
    } \\
  \subfloat[]{
    \includegraphics[scale=0.3]{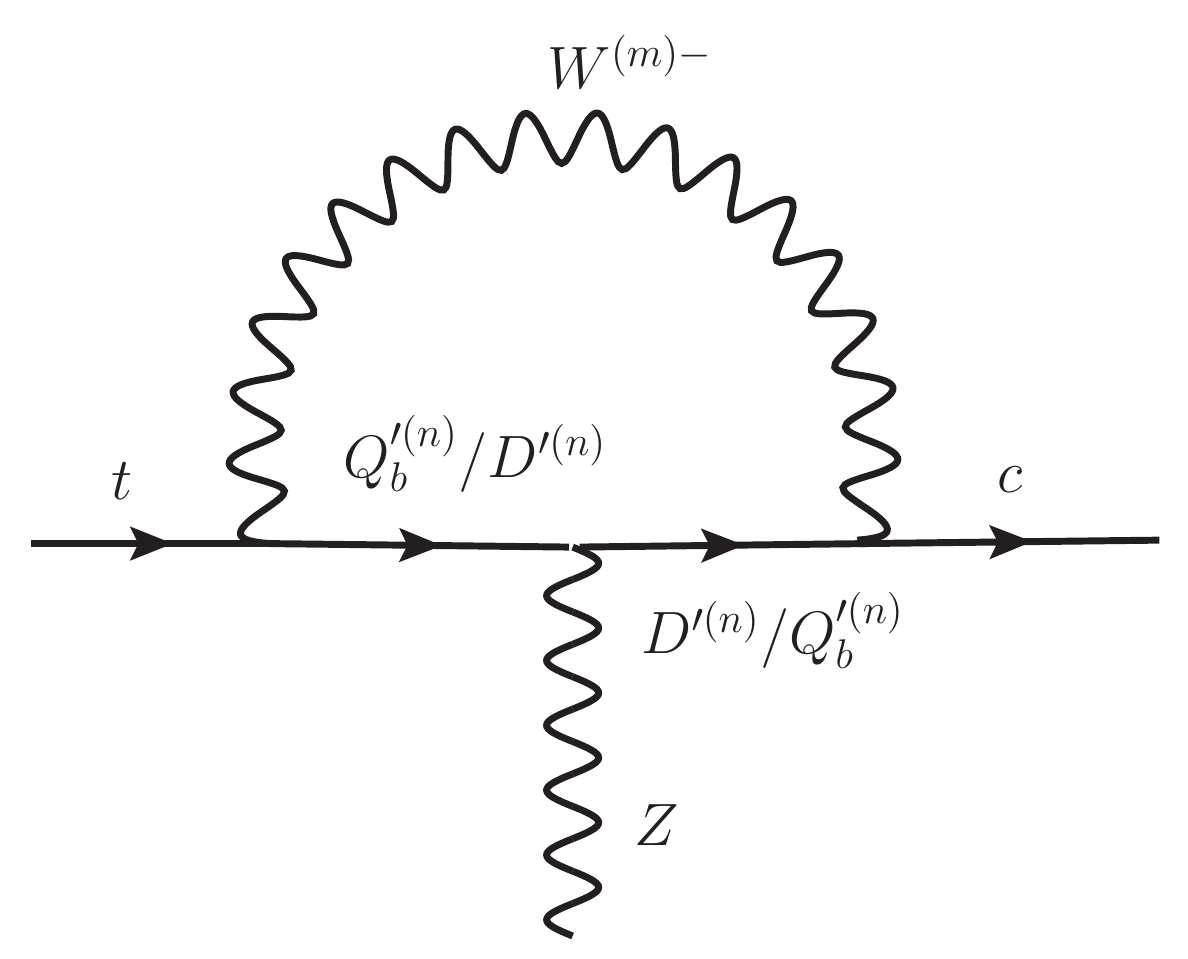}
    }    
  \subfloat[]{
    \includegraphics[scale=0.3]{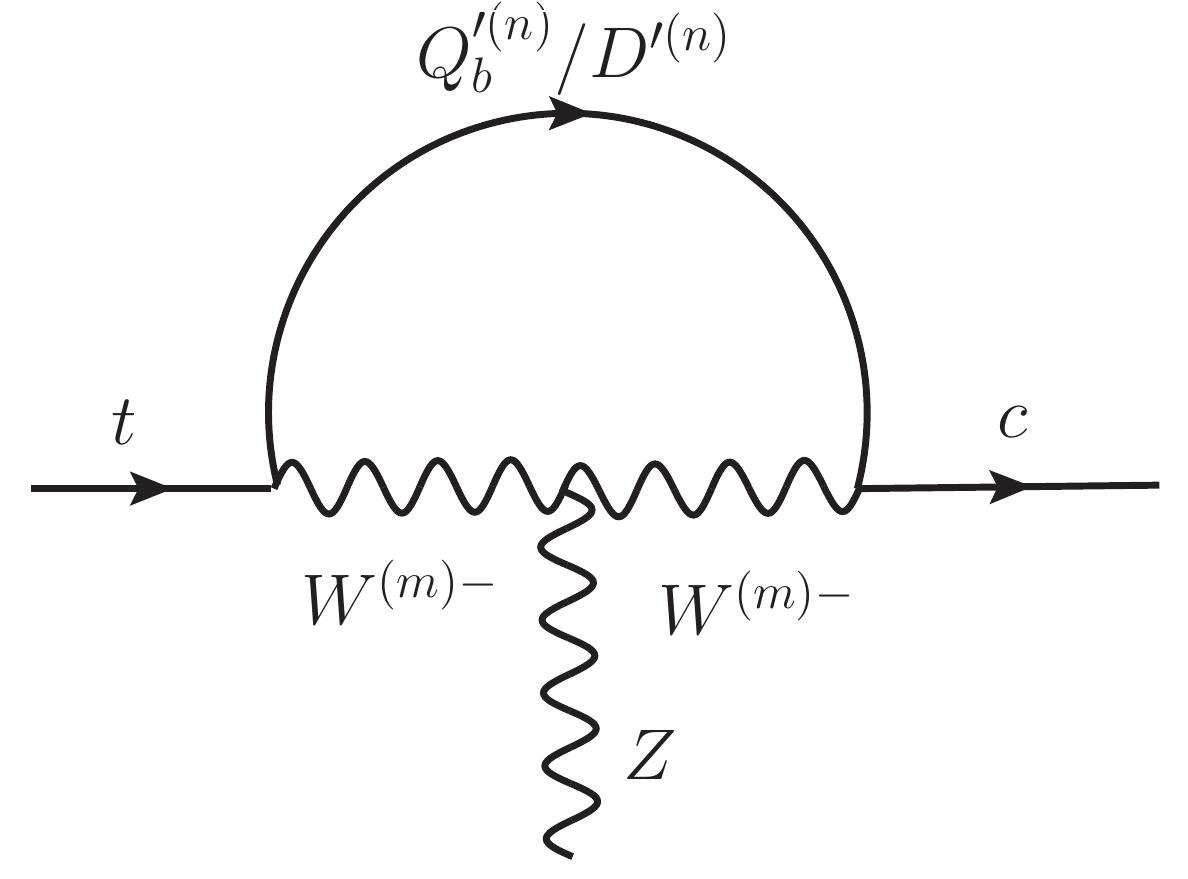}
    }
  \subfloat[]{
    \includegraphics[scale=0.3]{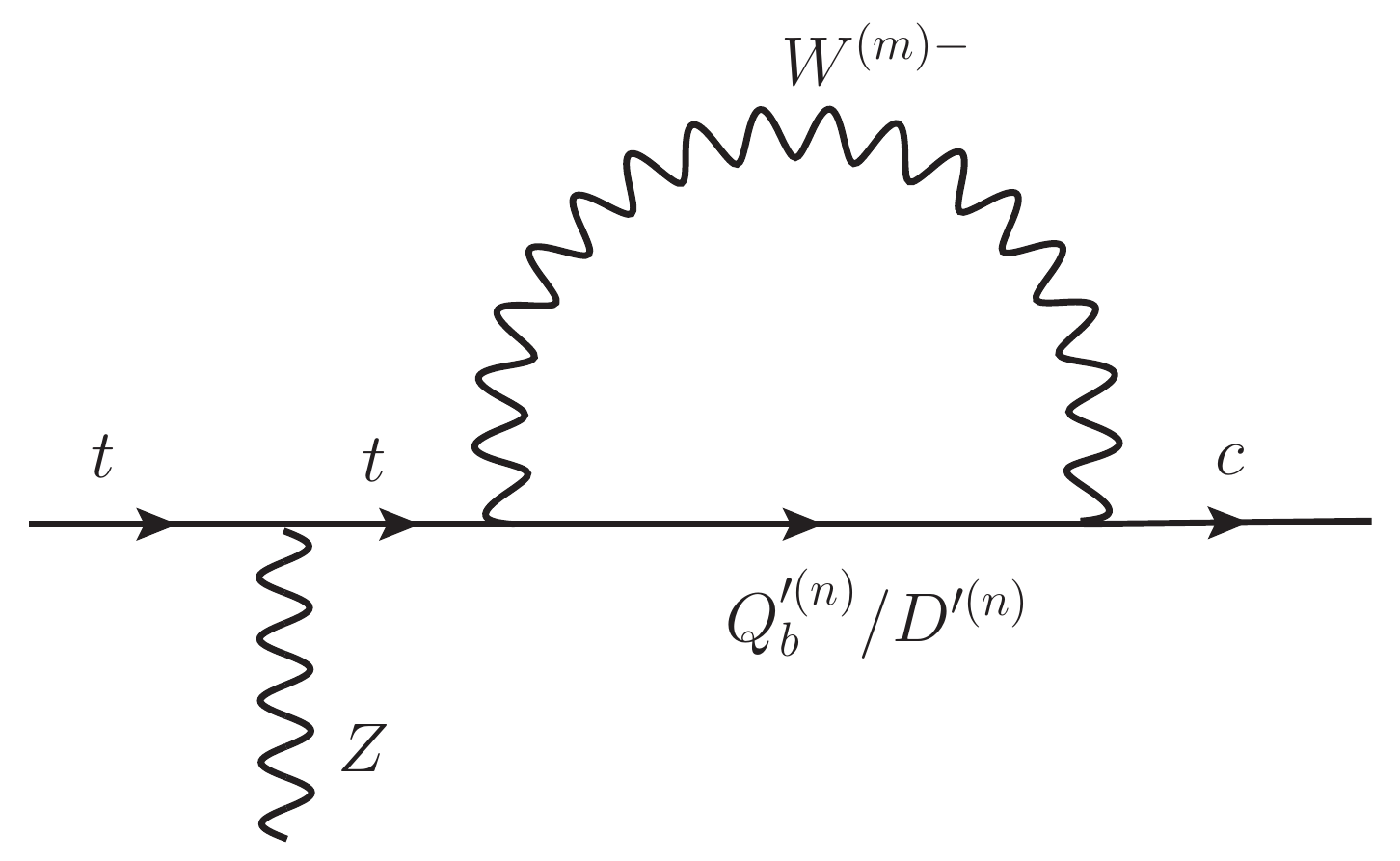}
    } \\
   \subfloat[]{
    \includegraphics[scale=0.3]{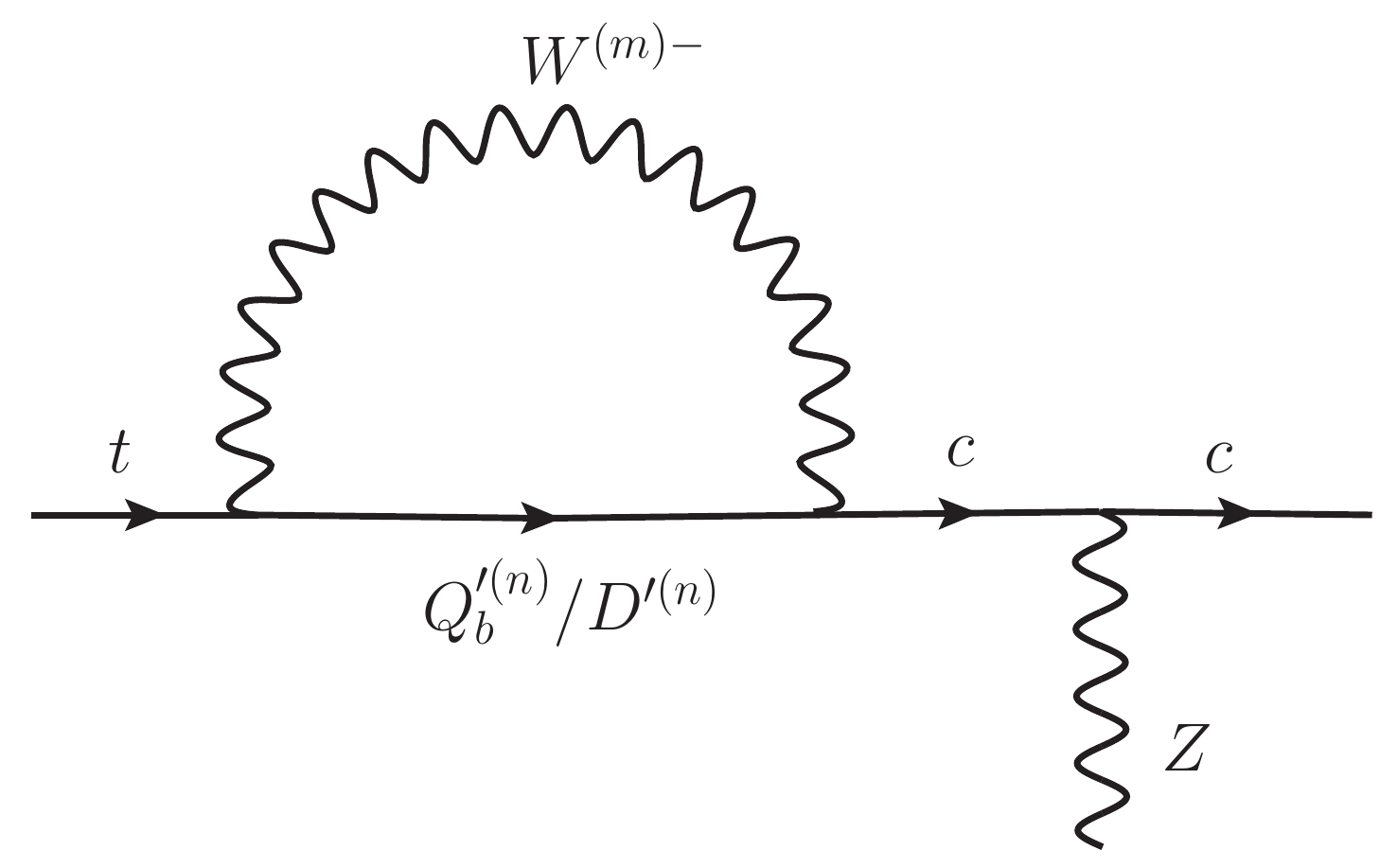}
    } 
    \caption[]{Feynman diagrams for the process $t\to cZ$ in the 't Hooft--Feynman gauge in the nmUED model. Though not explicitly shown, the diagrams with vertices of KK numbers $00m$ allowed by the KK parity are also included in our numerical calculations.}
\label{fig:tcZ}
\end{figure}
The most general decay width obtained from the above amplitude can be written as
\begin{align}
\Gamma_{t\to cZ} = \frac{1}{32\pi m_{t}^{3}}
                   &\sqrt{\{m_{t}^{2}-(m_{Z}-m_{c})^{2}\}
                     \{m_{t}^{2}-(m_{Z}+m_{c})^{2}\}} \notag \\
                   & \times \bigg[\mathscr{A}
                      \left(|A_{L}|^{2} + |B_{R}|^{2}\right) + 
                      \mathscr{B}
                        \left(|\tilde{A}_{L}|^{2} 
                              + |\tilde{B}_{R}|^{2}\right)
                    + \mathscr{C}{\rm Re} 
                     \left(A_{L} \tilde{B}_{R}^{\ast} + 
                           B_{R}^{\ast} \tilde{A}_{L}\right)
                           \notag \\
                   & \qquad + \mathscr{D}{\rm Re}  
                      \left(A_{L} B_{R}^{\ast}\right)
                    + \mathscr{E}{\rm Re}
                      \left(\tilde{A}_{L} 
                      \tilde{B}_{R}^{\ast}\right)      
                      \bigg],
\end{align}
where,
\begin{gather}
\mathscr{A} = m_{t}^{2} + m_{c}^{2} - 2m_{Z}^{2} 
              + \frac{(m_{t}^{2} - m_{c}^{2})^{2}}
                 {m_{Z}^{2}}, \; \;
\mathscr{B} = \frac{2(m_{t}^{2} - m_{c}^{2})^{2} 
                    - m_{Z}^{2}(m_{t}^{2} + m_{c}^{2} 
                    + m_{Z}^{2})}
                    {(m_{t} + m_{c})^{2}} \;, \\
\mathscr{C} = \frac{6m_{t}(m_{c}^{2} - m_{t}^{2} 
                     + m_{Z}^{2})}
                    {m_{t} + m_{c}}, \; \;
\mathscr{D} = - 12 m_{t} m_{c}, \; \; 
\mathscr{E} = - \frac{12m_{t}m_{c}m_{Z}^{2}}
                      {(m_{t} + m_{c})^{2}} \;.
\end{gather}
The relevant Feynman diagrams for the $t\to cZ$ decay are shown in Fig.~\ref{fig:tcZ}. The remarks on KK indices made in Section~\ref{sbsc:tcg} also apply to the evaluation of $A_L$ and $\tilde B_R$.

\section{Results}
\label{sec:results}

Before presenting the results of the $t\to cg$ and $t\to cZ$ decays in (n)mUED, we quickly review the status of the SM expectations on these decays. The dominant decay mode of the top quark is $t\to bW$, the decay width of which is given by
\footnote{We note and correct a typo in the corresponding formula given in ~\cite{Dey:2016cve}.}
%
%
%
\begin{align}
\Gamma_{t\to bW} = \frac{G_{F}}{8\sqrt{2}\pi}|V_{tb}|^{2}
                    m_{t}^{3}
                    \left[
                    1 - 3\left(\frac{m_{W}}{m_{t}}\right)^{4} 
                      + 2\left(\frac{m_{W}}{m_{t}}\right)^{6}                       
                    \right].
\end{align}
For $G_{F} = 1.166\times 10^{-5}$ GeV$^{-2}$, $m_{W} = 80.39$~GeV, $m_{t} = 174.98$~GeV~\cite{Abazov:2014dpa}, $\Gamma_{t\to bW} \sim 1.5$~GeV. 
Since this is the most prominent decay mode of the top quark, the branching ratio of any other mode $t\to X$ is virtually given by
\begin{align}
\text{BR}(t\to X) = \frac{\Gamma_{t\to X}}{\Gamma_{t\to bW}}.
\end{align}
The SM prediction for the $t\to cg$ branching ratio
\footnote{Aside from a small increase in the phase space, the $t\to u$ conversion will be even more suppressed than the $t\to c$ conversion by a factor of $|V_{ub}|^{2}/|V_{cb}|^{2}\sim 10^{-4}$.}
is
$
\text{BR}(t\to cg) = \left(4.6^{+1.1}_{-0.9} 
                     \pm 0.4^{+2.1}_{-0.7}
                     \right)\times 10^{-12},
$
where the first uncertainty is due to the bottom mass uncertainties, the second from the CKM mixing angle uncertainties, and the third due to the variation in the renormalization scale between $m_{Z}$ (+ sign) and $1.5m_{t}$ ($-$ sign)~\cite{AguilarSaavedra:2002ns}. Also, according to Ref.~\cite{Eilam:1990zc}, where a pole mass of $m_{b} = 5$~GeV has been used, the branching ratio is ten times higher than this value. Taking the pole mass of the $b$-quark as $4.18$~GeV~\cite{Patrignani:2016xqp}, our SM prediction for BR$(t\to cg)$ is $2.42\times 10^{-11}$. If we take the running mass $\bar{m}_{b}(m_{t}) = 2.74$~GeV, the branching ratio is $5.54\times 10^{-12}$. It is worth noting that these branching ratios are more sensitive to the bottom quark mass than the top quark mass as the leading $m_{t}$ dependence cancels in the branching ratio.

The branching ratio of $t\to cZ$ in the SM is estimated to be $(1.03\pm 0.06)\times 10^{-14}$ by~\cite{Abbas:2015cua}, which is again one order of magnitude smaller than the result of~\cite{Eilam:1990zc} due to the different values of input parameters. For $m_{Z} = 91.19$~GeV, our SM prediction for the $t\to cZ$ branching ratio is $6.5\times 10^{-14}$ when taking the pole mass of the $b$-quark to be $4.18$~GeV and it becomes $1.19\times 10^{-14}$ when using the running mass $\bar{m}_{b}(m_{t}) = 2.74$~GeV. As in the case of $t\to cg$, this decay branching ratio is also more sensitive to the bottom quark mass.

\subsection{The $t\to cg$ Decay}
\label{sbsc:tcgres}

\subsubsection{mUED Result}

In the mUED scenario, the loop-induced $t\to cg$ decay gets additional contributions from the higher KK particles running in the loop. The relevant Feynman diagrams are already shown in Fig.~\ref{fig:tcg}. Due to the conservation of KK number in mUED, the KK indices $m$ and $n$ in each vertex of the diagrams should be equal. The difference from the SM in mUED is basically the presence of KK counterparts of SM particles in the loop as well as the presence of charged KK scalars. Also, the mixing in the KK fermion sector plays an important role, as discussed in Section~\ref{sbsc:Physcl_Eigenst}. We note in passing that in mUED, the only relevant parameter is the inverse of compactification radius $1/R$ and the masses of all the KK particles solely depend on this quantity.

\begin{figure}[t]
\begin{center}
\includegraphics[scale=0.6]{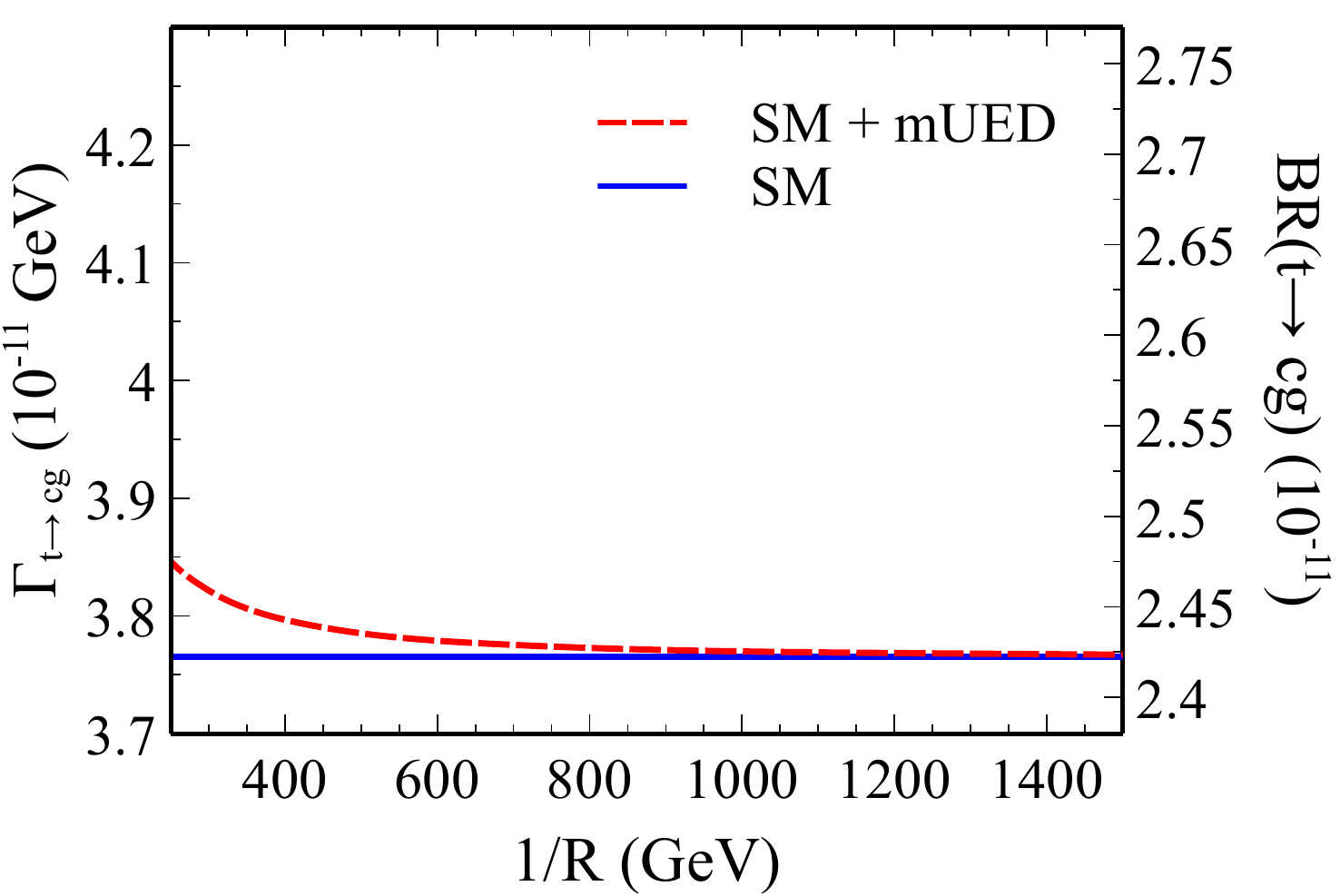}
\caption{The decay width of the process $t\to cg$ as a function of the inverse compactification radius $1/R$ in the case of mUED.}
\label{f:tcg_mued}
\end{center}
\end{figure}

Fig.~\ref{f:tcg_mued} shows the dependence of $t\to cg$ decay width and branching ratio on $1/R$.  The blue solid horizontal line represents the SM expectation, and the red dashed curve represents the result when the mUED contribution is taken along with the SM. The tail end of the red dashed curve at higher values of $1/R$ approaches the SM result, clearly showing the decoupling behavior of the KK mode contribution. Our result is consistent with that in Ref.~\cite{Aranda:2013xra}, where the $t\to cg$ decay has been considered in the mUED setup but includes only the first KK mode contribution. From this it is evident that even for the lower values of $1/R$, the magnitude of the decay width does not change much. Therefore, within the mUED scenario it is impossible to enhance the branching ratio of $t\to cg$ to any significant level, not to mention that the lower values of $1/R$ are already ruled out by the LHC data; the recent studies including LHC data exclude $1/R$ up to 1.4~TeV~\cite{Choudhury:2016tff, Beuria:2017jez, Chakraborty:2017kjq, Deutschmann:2017bth}. 

\subsubsection{nmUED Results}

As alluded to earlier, the presence of BLKTs modifies the spectrum and couplings of the KK modes from the mUED scenario. The BLKT parameters determine the mass spectrum and the overlap integrals, which are functions of BLKT parameters and control the couplings. The loop-induced $t\to cg$ decay in this case differs from the mUED case in the following aspects. Firstly, since in nmUED the KK parity rather than the KK number is a good quantum number, one can have tree-level 0-0-$n$ couplings for even $n$.  This gives rise to a few more Feynman diagrams that contribute to the process, {\it i.e.}, in Fig.~\ref{fig:tcg}, diagrams with appropriate $m~(\mbox{or}~n)=0$ also contribute. Secondly, there are modifications in the KK masses and couplings caused by the BLKT parameters, the choices and ranges of which are given in Section~\ref{s:Lag_nmUED}. We have universal BLKT parameters $R_{\Phi}$ and $R_{f}$ for bosons and fermions \footnote{Since the BLKT parameters are dimensionful parameters, we use the dimensionless quantity $R_X = r_X /R$ when presenting our results.}, respectively, and they are assumed to be positive-definite. With these conditions, we consider two possibilities: (i) $R_{\Phi} = R_{f}$ and (ii) $R_{\Phi}\neq R_{f}$.  

%
\begin{figure}[t]
\begin{center}
\includegraphics[scale=0.6]{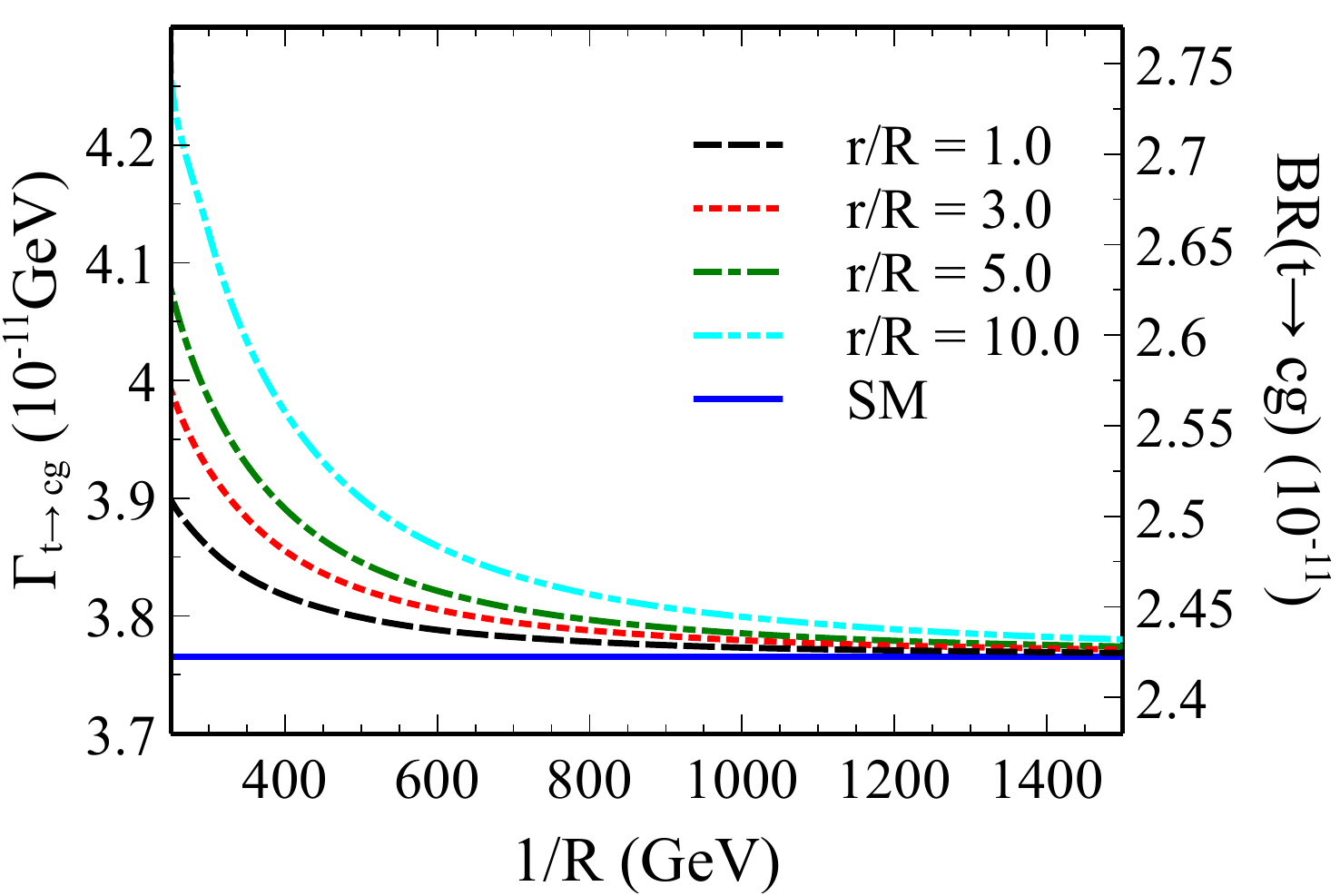}
\caption{Decay width and branching ratio for the $t\to cg$ process as a function of $1/R$ in the nmUED scenario for different BLKT parameters. In this case we consider a universal BLKT parameter $r = r_\Phi$.}
\label{f:tcg_nmuedsame}
\end{center}
\end{figure}

We first consider case (i), $R_{\Phi} = R_{f} \equiv r/R$, {\it i.e.}, we have a universal BLKT parameter in the model. Thanks to the orthogonality conditions in this scenario, the overlap integrals that modify the couplings will become unity and thus the scenario \textit{almost} reduces to the mUED case. The couplings remain the same as that of the mUED but the KK mases are now dependent on the common BLKT parameter $r/R$ via a transcendental equation. Fig.~\ref{f:tcg_nmuedsame} shows the decay width and branching ratio for different values of the universal BLKT parameter. A quick glance at Figs.~\ref{f:tcg_nmuedsame} and \ref{f:tcZ_nmuedsame} reveals that for non-zero universal BLKT cases, the decay width grows with $r$ for fixed $1/R$. This is more prominent in the lower $1/R$ region. The reason for this is that with the increase in the value of BLKT parameter, the KK masses decrease, making the propagator less suppressed in the nmUED case.
%

\begin{figure}[H]
\begin{center}
\subfloat[\label{sf:tcgrp8}]{
\includegraphics[scale=0.5]{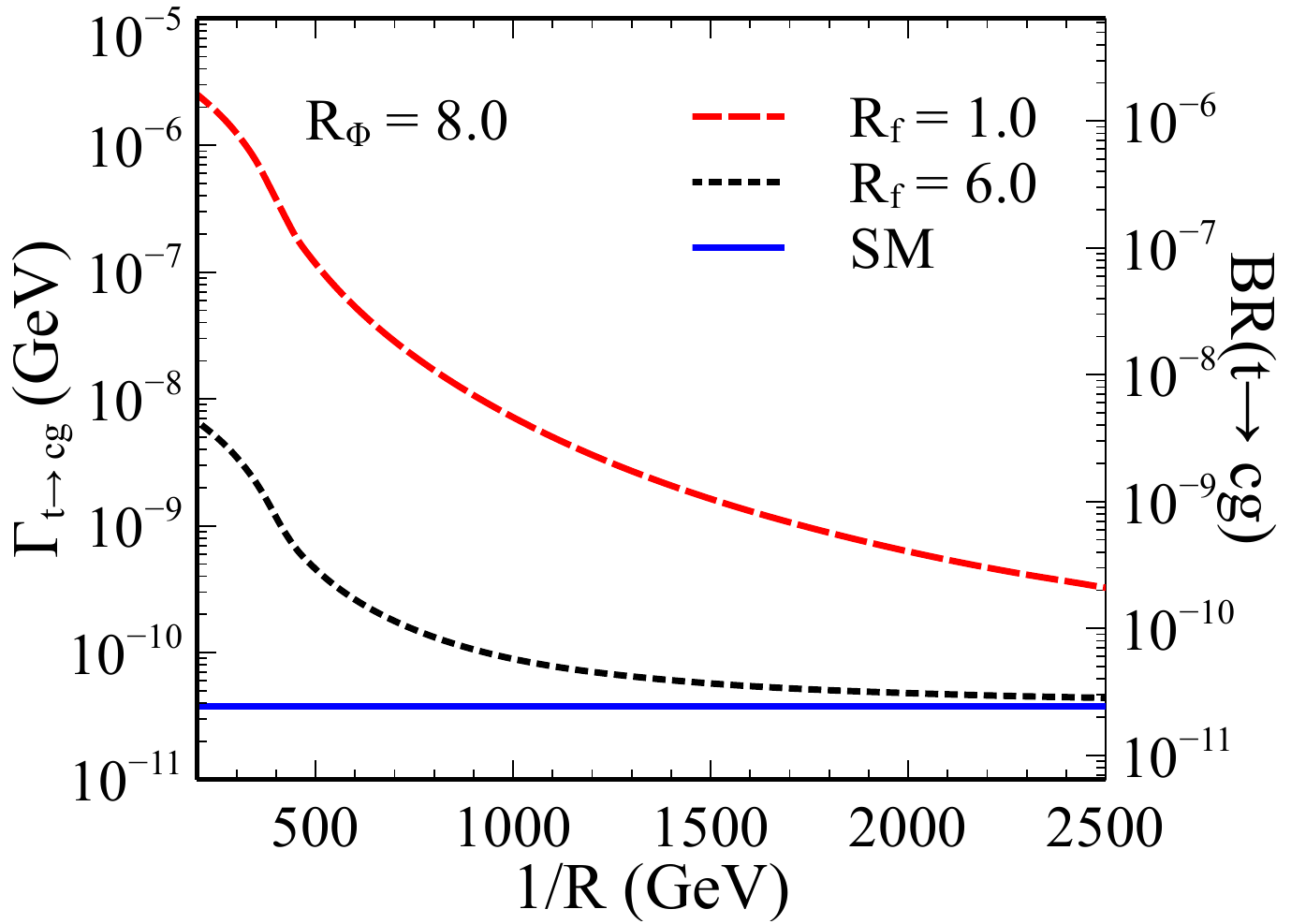}}
~~~~
\subfloat[\label{sf:tcgrqp1}]{
\includegraphics[scale=0.5]{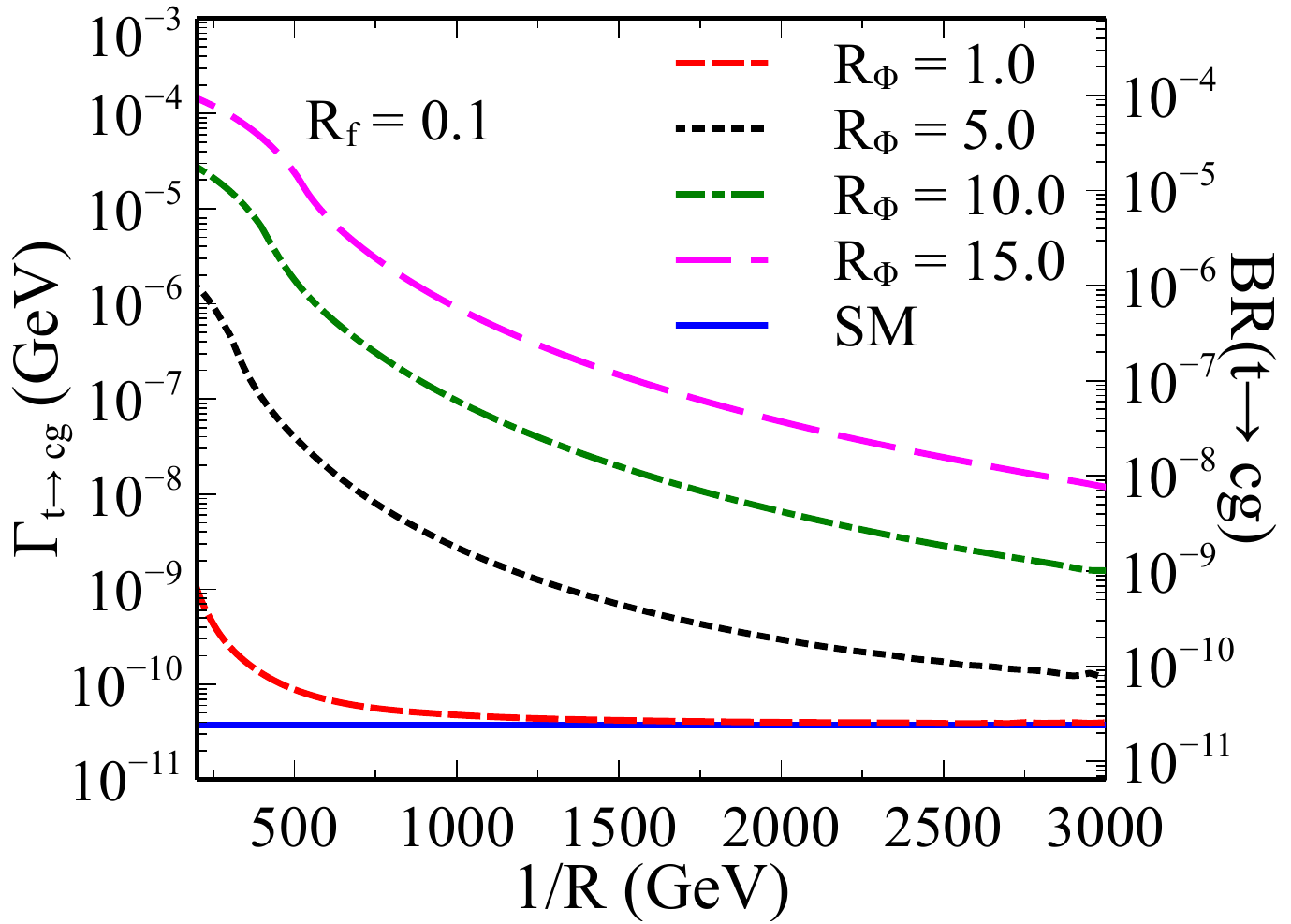}} 
\\
\subfloat[\label{sf:tcgrq3}]{
\includegraphics[scale=0.5]{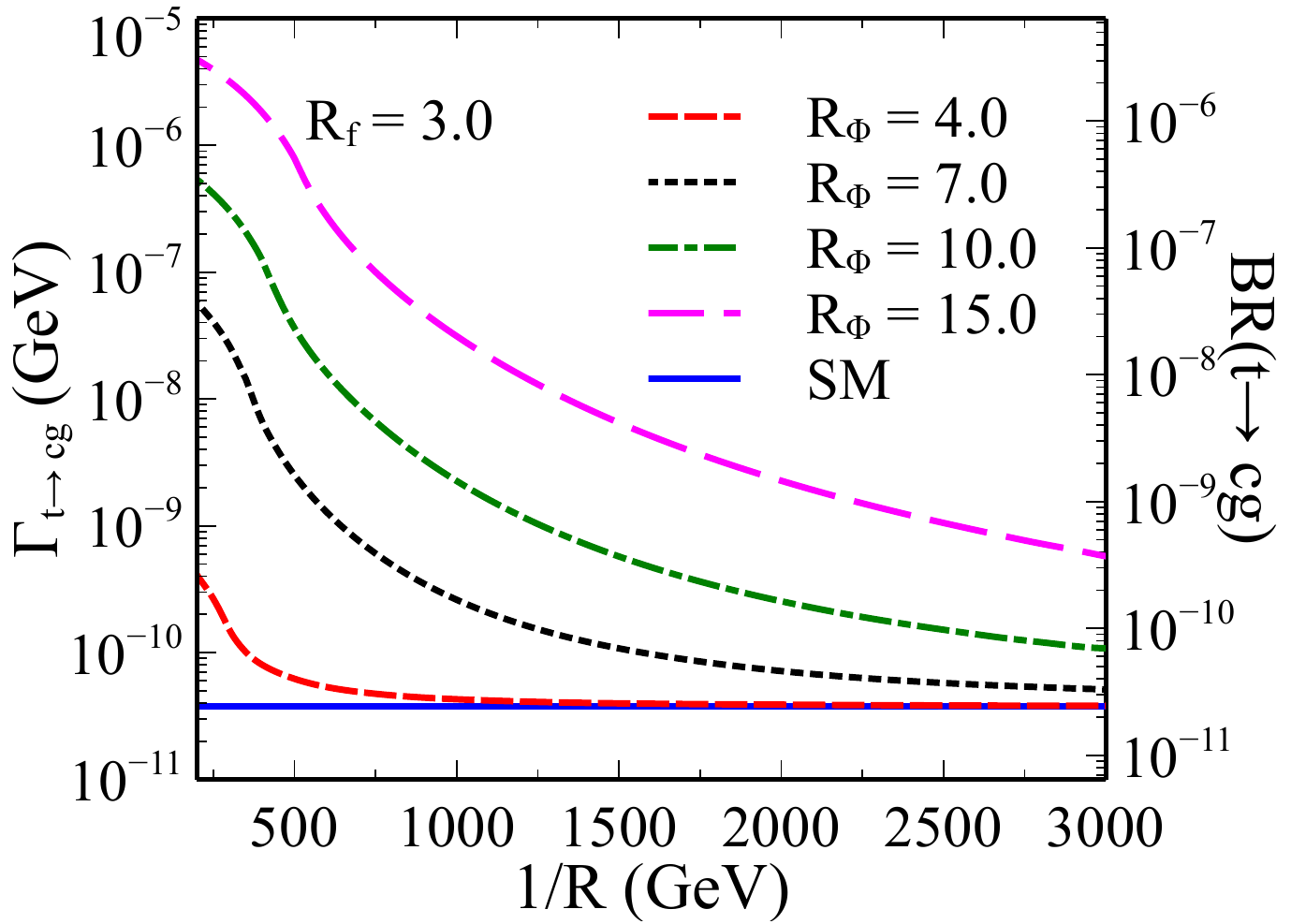}}
~~~~
\subfloat[\label{sf:tcgrq7}]{
\includegraphics[scale=0.5]{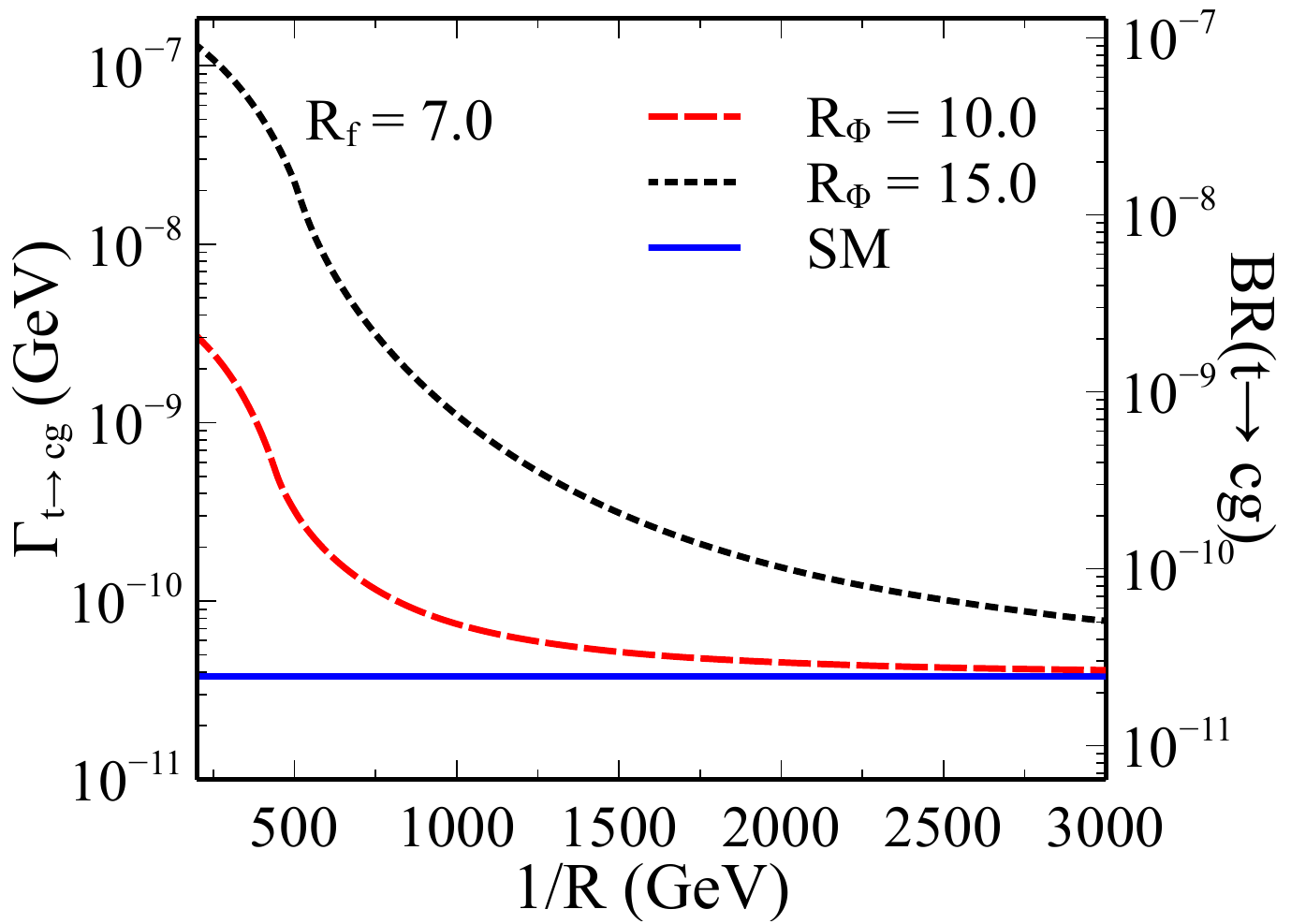}}
\caption{Decay width and branching ratio for the $t\to cg$ decay as a function of $1/R$ in the nmUED scenario for different values of the BLKT parameters $R_{\Phi}~(=r_{\Phi}/R)$ and $R_{f}~(=r_{f}/R)$.}
\label{f:tcg_nmueddiff}
\end{center}
\end{figure}

In case (ii), $R_{\Phi} \neq R_{f}$, {\it i.e.}, we have one universal BLKT parameter for the bosonic sector and another for the fermionic sector. This brings in two new features. Firstly, this implies that depending on their respective BLKT parameters, the KK excitations of fermions and the corresponding KK scalar/gauge bosons have different masses. Secondly, the couplings of KK particles are now modulated by the overlap integrals of the form given in Eq.~\eqref{eq:oi}. These overlap integrals in turn depend on the KK masses and, in some cases, on the mass difference between the two different types of KK particles ({\it e.g.}, some overlap integrals depend on the difference between $M_{\Phi_n}$ and $M_{Q_n}$). One important point to note in the case (ii) is that we always make the choice $R_{\Phi} > R_{f}$. This is because the KK masses decrease with increasing values of BLKT parameters. We need $R_{\Phi} > R_{f}$ to ensure that none of the KK fermions in this KK parity-conserving scenario become the lightest KK particle (LKP) so as to claim the candidacy of DM. The choice $R_{\Phi} > R_{f}$ maintains the first level KK photon as the DM candidate. For the DM aspects of the nmUED scenario, see Refs.~\cite{Datta:2013nua, Flacke:2017xsv, Dasgupta:2018nzt}. In Fig.~\ref{f:tcg_nmueddiff}, we show the variation of $t\to cg$ decay width for various choices of BLKT parameters. According to these plots, the decay width can be a few orders of magnitude larger than the SM expectation for some choices of the BLKT parameters. Clearly, the effect is greater in the lower $1/R$ region than the higher $1/R$ region. But lower $1/R$ values suffer from collider and other constraints which we will discuss in Section~\ref{sec:constrnts}. The difference in the $t\to cg$ decay width between the case with a universal BLKT parameter and the case with different BLKT parameters is precisely due to the two features alluded to earlier. This is why a simple-minded interpolation from the latter to the former is impossible.
We would like to mention that the assumption of universality in the bosonic and fermionic BLKT parameters can be relaxed and will lead to a richer structure. Although a rigorous quantitative estimate would be beyond the scope of the present work, qualitatively, it can be understood as follows. First, note that if we relax the universal fermionic BLKT set-up and move to the generation-wise fermionic BLKT parameters then we may have, for example, three fermionic BLKT parameters which will determine the relevant overlap integrals. Clearly there will be more freedom to tune those integrals which actually modulate the relevant couplings. Secondly, if we take different BLKT parameters for scalars and gauge bosons a similar effect will ensue. Thus, the decay widths can be larger even for higher values of the inverse compactification radius $1/R$ for non-universal BLKTs.

\subsection{The $t\to cZ$ Decay}
\label{sbsc:tcZres}

\subsubsection{mUED Result}

The Feynman diagrams for the $t\to cZ$ decay is shown in Fig.~\ref{fig:tcZ}. Like the previous $t\to cg$ case, the KK indices $m$ and $n$ have to be equal in mUED due to the conservation of KK number.

\begin{figure}[H]
\begin{center}
\includegraphics[scale=0.6]{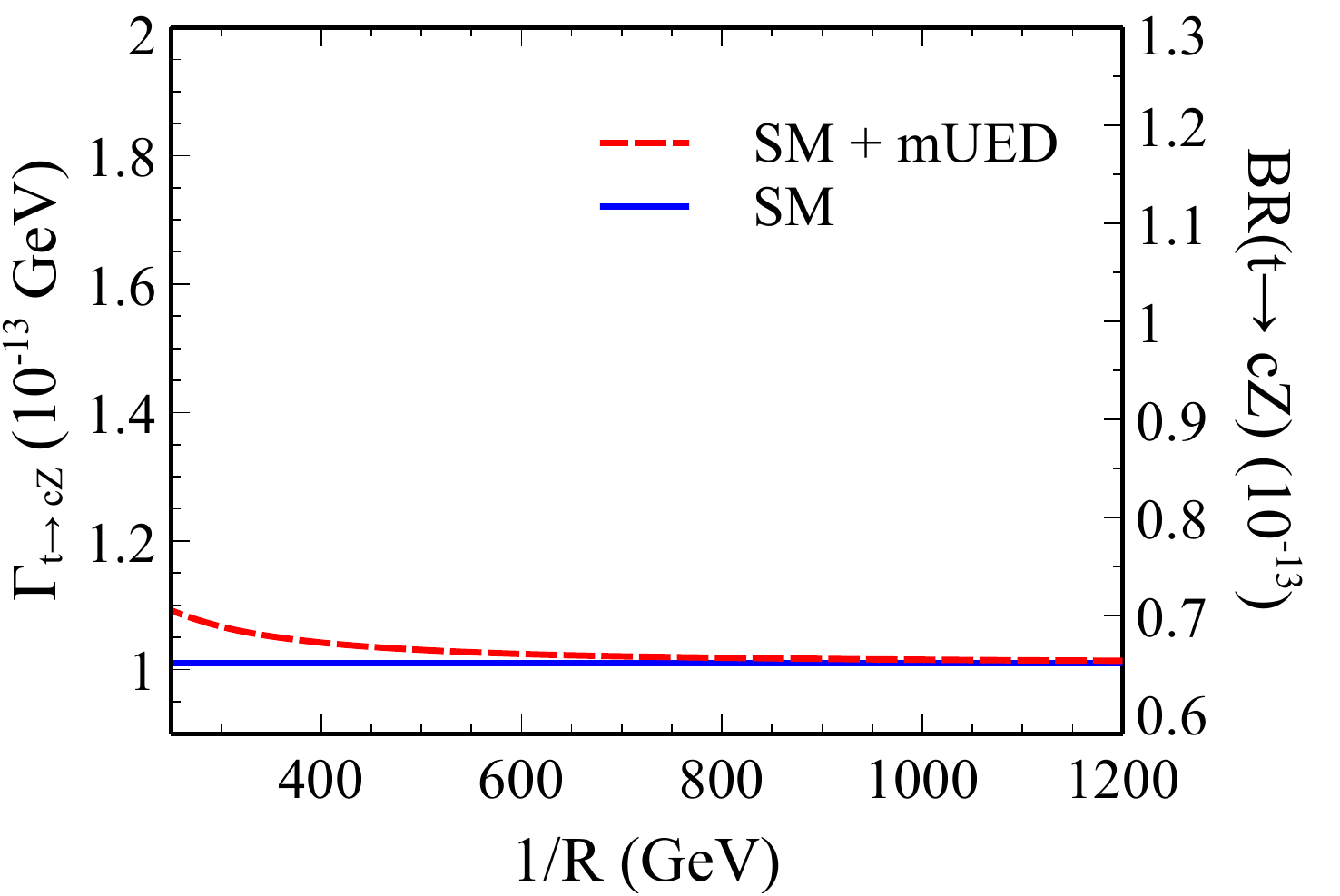}
\caption{\small Decay width and branching of the $t\to cZ$ process as a function of the inverse compactification radius $1/R$ in the case of mUED.}
\label{f:tcZ_mued}
\end{center}
\end{figure}

In Fig.~\ref{f:tcZ_mued}, we show how the decay width vary with $1/R$.  The blue solid horizontal line gives the SM value, and the red dashed curve represents the mUED result. Similar to the $t\to cg$ decay, here we also find no significant enhancement in the decay width for allowed values of $1/R$.

\subsubsection{nmUED Results}

\begin{figure}[H]
\begin{center}
\includegraphics[scale=0.55]{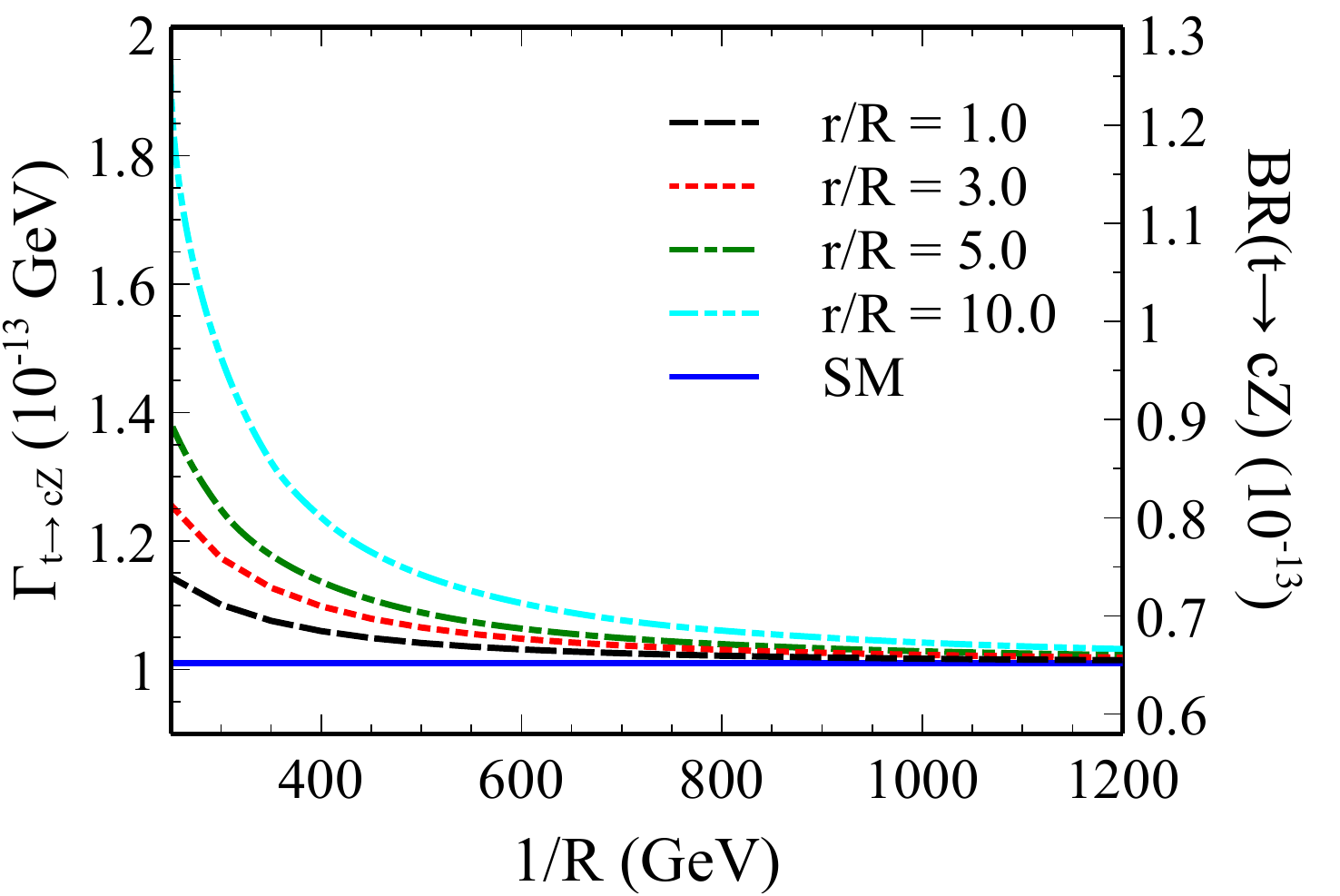}
\caption{Decay width and branching ratio of the $t\to cZ$ process as a function of $1/R$ in the case of nmUED for different values of the universal BLKT parameter $r$.}
\label{f:tcZ_nmuedsame}
\end{center}
\end{figure}

Analogous to the $t\to cg$ case, here we also present the results of two cases: (i) $R_{\Phi} = R_{f}$ and (ii) $R_{\Phi} \neq R_{f}$ sequentially. Fig.~\ref{f:tcZ_nmuedsame} shows the variation of the $t\to cZ$ decay width in the universal BLKT scenario, $R_{\Phi} = R_{f} \equiv r/R$, for different values of $r/R$. Clearly, there is no significant enhancement in the width for higher values of $1/R$ which may be insulated from the current LHC searches.

%

\begin{figure}[H]
\begin{center}
\subfloat[\label{f:rp8}]{
\includegraphics[scale=0.35]{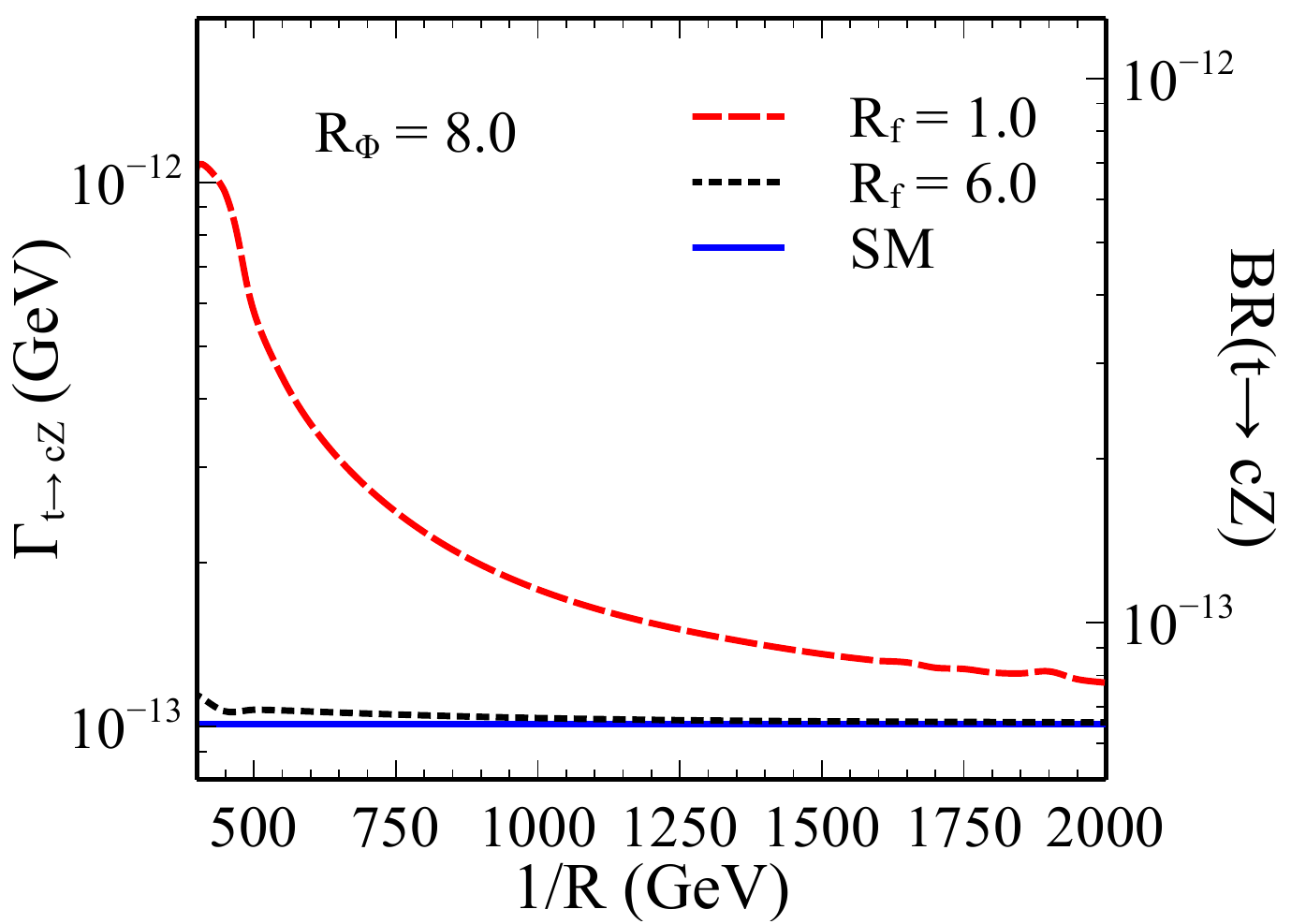}} ~
\subfloat[\label{f:rqp1}]{
\includegraphics[scale=0.35]{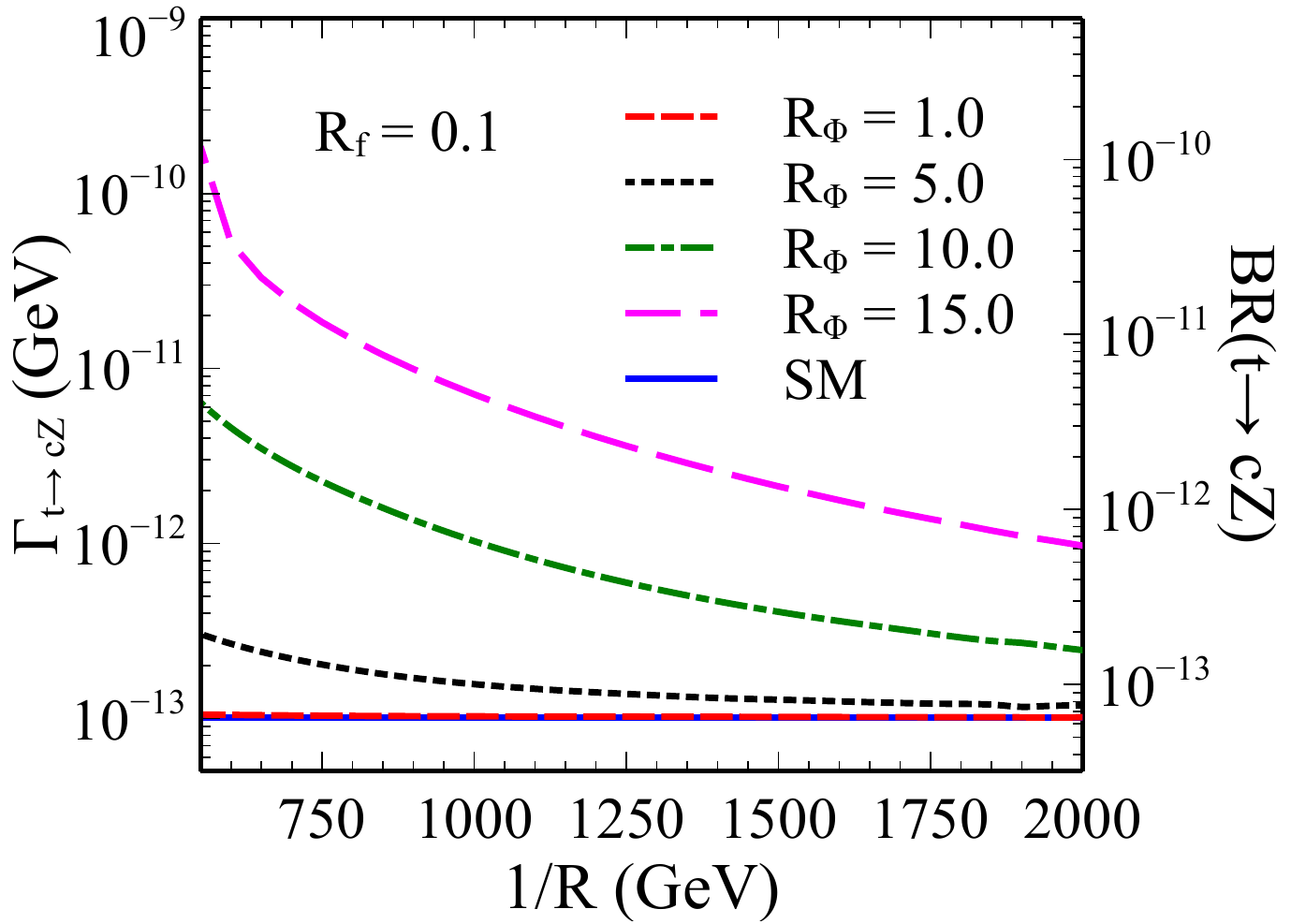}} ~
\subfloat[\label{f:rq3}]{
\includegraphics[scale=0.35]{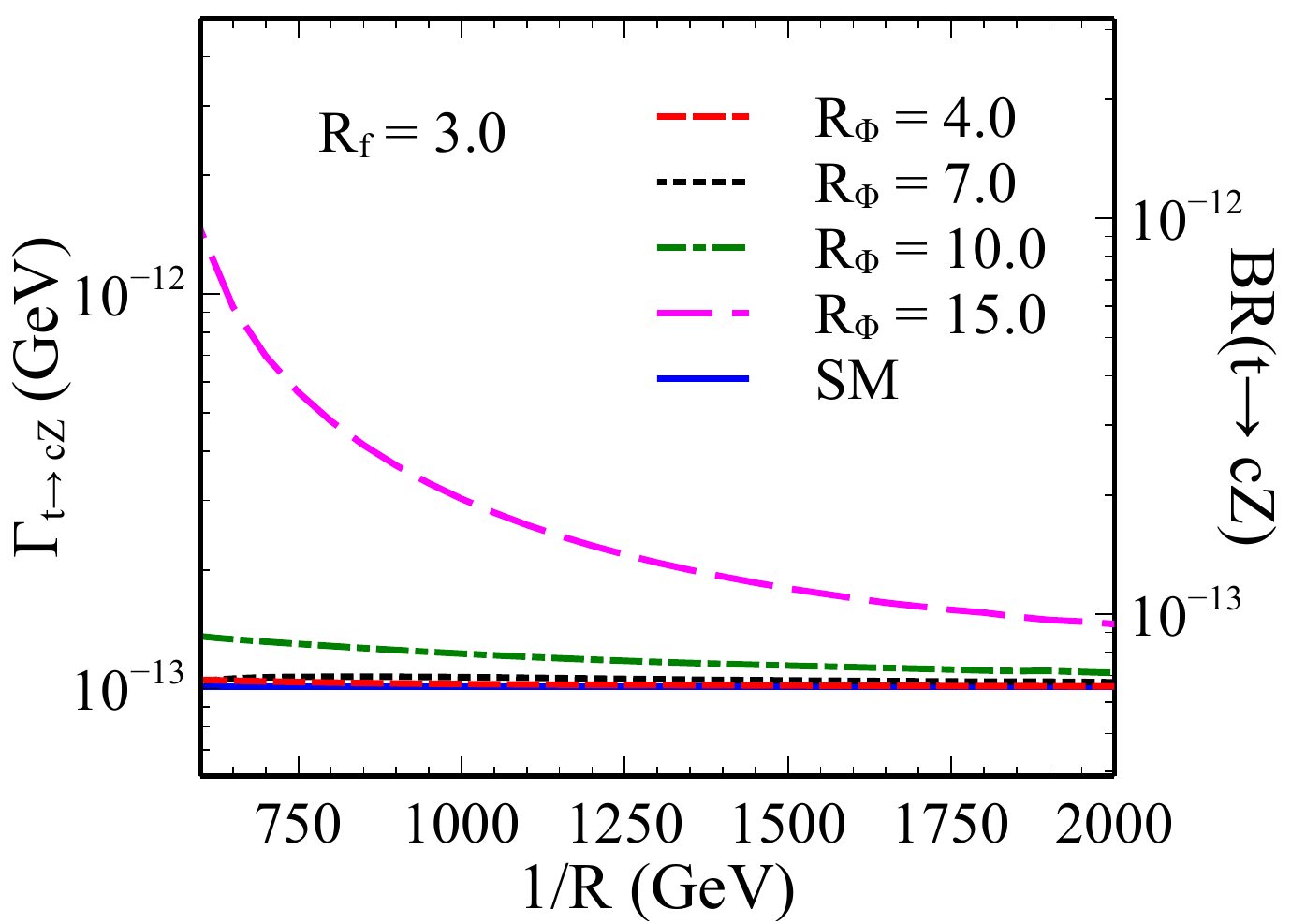}}
\caption{The decay width for the process $t\to cZ$ as a function of $1/R$ in the case of nmUED for different values of BLKT parameters $R_{\Phi}~(=r_{\Phi}/R)$ and $R_{f}~(=r_{f}/R)$.}
\label{f:tcZ_nmueddiff}
\end{center}
\end{figure}

Almost similar is the scenario when we take case (ii), as is evident by a comparison between Fig.~\ref{f:tcZ_nmuedsame} and Fig.~\ref{f:tcZ_nmueddiff}. However, in this case for very small $R_{f}$ and $1/R~\sim$ 2 TeV the width can be at least one order of magnitude higher than the SM as can be seen from the Fig.~\ref{f:rqp1}.  

\section{Precision Data and LHC Phenomenology}
\label{sec:constrnts}

\begin{figure}[t]
\begin{center}
\includegraphics[scale=0.9]{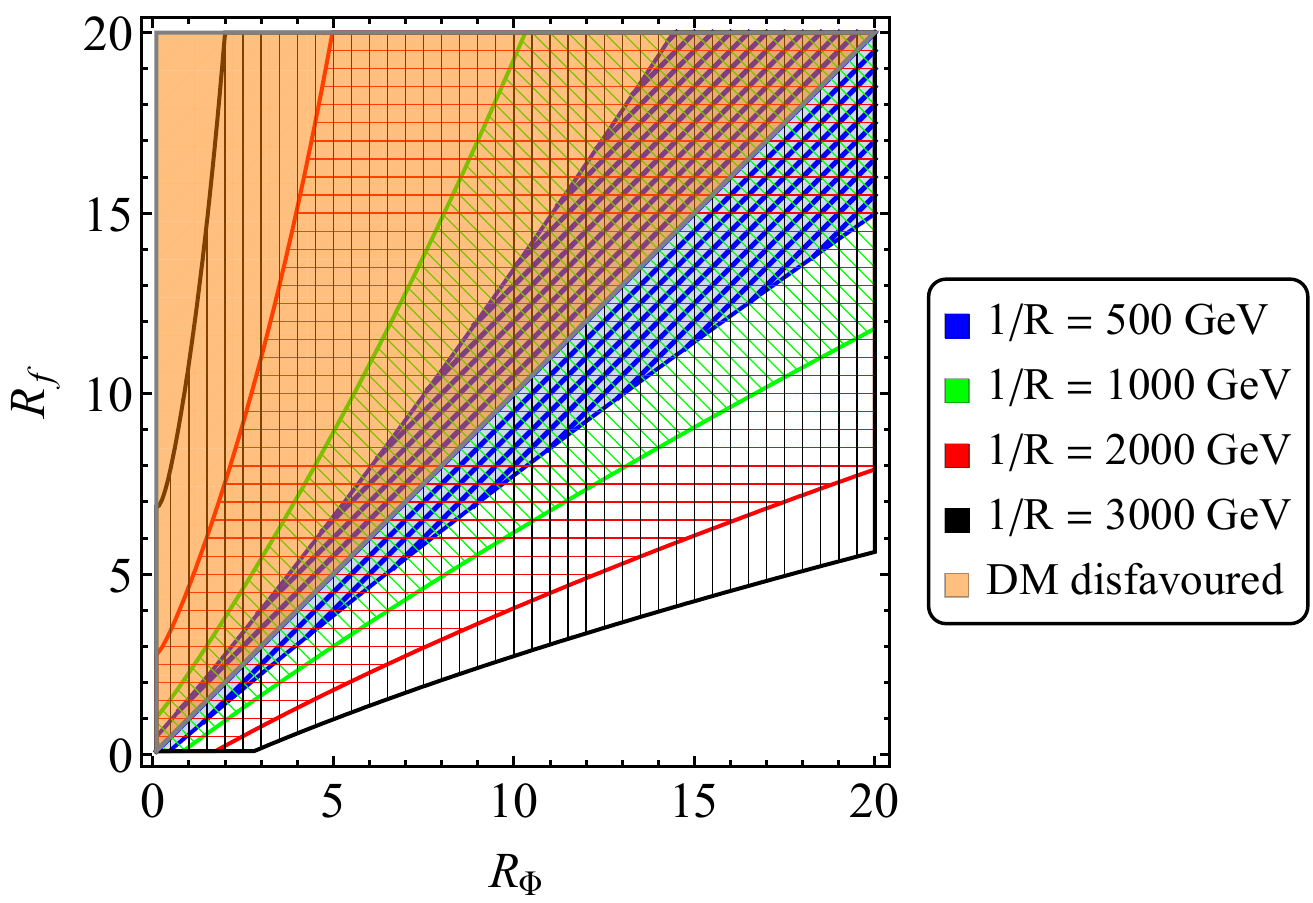}
\caption{Regions in the $R_{\Phi}-R_{f}$ plane allowed by the electroweak precision data at $2\sigma$ level for different choices of $1/R$. The orange shaded upper half region is disfavored because in that case the model would have a fermionic LKP.}
\label{f:stu}
\end{center}
\end{figure}

In this section, we discuss the issues of electroweak precision data ({\it i.e.}, the $S,T,U$ parameters) and the LHC data. The aim here is to examine how well the ranges of BLKT parameters, that give enhancements in the decay widths, fare under the constraints from precision data as well as LHC observations.   
\noindent
{\bf Precision data:} In Ref.~\cite{Dey:2016cve}, we have discussed the constraints on the BLKT parameters coming from the electroweak precision data. Using the same methodology there, we show in the $R_{\Phi}-R_{f}$ plane the region consistent with the precision data at $2\sigma$ level for different values of $1/R$ in Fig.~\ref{f:stu}. The orange shaded upper diagonal region, for which $R_{f} > R_{\Phi}$, is disfavored from the DM point of view as this region corresponds to the scenario where a KK fermion becomes the LKP. A combined study of Fig.~\ref{f:tcg_nmueddiff} and Fig.~\ref{f:stu} shows that the parameters which give rise to a larger enhancement in the $t\to cg$ decay width are in tension with the electroweak precision data. To give an example, one can have in Fig.~\ref{sf:tcgrqp1} almost three orders of magnitude enhancement in $\Gamma_{t\to cg}$ compared to the SM for $1/R = 3000$~GeV, $R_{f} = 0.1$ and $R_{\Phi} = 15$. But this set of parameters is clearly not within the allowed range of the precision data, as shown in Fig.~\ref{f:stu}. At this point, it is tantalizing to argue that lower values of $1/R$ can give even larger enhancements in $\Gamma_{t\to cg}$ while still being within the allowed range of the precision data. But the main hindrance in this assertion comes from the LHC observations, {\it e.g.}, the dilepton data to be discussed next.

\noindent
{\bf LHC Phenomenology:} The second KK level gauge bosons can be singly produced at the LHC owing to the violation of KK number in the nmUED. The resonant production of these second KK level gauge bosons and their subsequent decays to SM fermions at the LHC, $pp\to X^{(2)}\to \bar{f}f,~(X = \gamma, Z)$ occur via the coupling $g_{X^{(2)}\bar{f}f}$. Now, using a dilepton final state, the LHC searches can put stringent constraints on the nmUED parameters. Recent ATLAS~\cite{ATLAS-CONF-2016-045} and CMS~\cite{CMS-PAS-EXO-16-031} searches for high-mass resonance in the dilepton channel using $\sqrt{s} = 13$~TeV data have been utilized to put bounds on the BLKT parameters~\cite{Flacke:2017xsv}. Without performing an extensive collider simulation, we simply translate and impose the bounds obtained in~\cite{Flacke:2017xsv} on the BLKT parameters of our interest. We compare the coupling $g_{X^{(2)}\bar{f}f}$ in our model with the constraint explicated in Ref.~\cite{Flacke:2017xsv} for a given $M_{X^{(2)}}$ to determine the exclusion region from LHC dilepton searches. Such a comparison shows that the parameters which can potentially give an enhanced $\Gamma_{t\to cg}$ decay is not allowed by the dilepton search data. Admittedly, a rigorous and proper analysis of the dilepton search constraints as well as monojet and other collider constraints (in the same vein as~\cite{Choudhury:2016tff, Beuria:2017jez, Chakraborty:2017kjq}) via a dedicated collider simulation is imperative, and we leave it to a future work. It is also interesting to note that relaxing the universality of the BLKT parameters for fermionic and bosonic sectors and considering the possibility of different BLKT parameters for different sets of fields can be useful in many aspects~\cite{Dasgupta:2018nzt}, and that will open up new possibilities in the rare decays.
One can get some upper bound on the BLKT parameters from e.g., unitarity constraints~\cite{Jha:2016sre}. However, we reiterate that although it is apparent from the previous plots that larger values of some of the BLKT parameters may lead to more enhanced branching ratios, it also has to be kept in mind that larger values reduce the masses of the KK particles for a fixed value of the compactification scale $1/R$ and this may have tension with LHC data.
Both ATLAS and CMS Collaborations have searched for these rare decays~\cite{lhcTopWGNov:2017}. Important constraints on the FCNC top decays come from the single top productions in the $pp$ collisions. Using the 20.3~fb$^{-1}$ data at $\sqrt{s} = 8$~TeV in search for single top production via FCNC processes, the ATLAS Collaboration sets an upper bound on the $t\to cg$ branching ratio as BR$(t\to cg) < 20 \times 10^{-5}$~\cite{Aad:2015gea}. Using combined $\sqrt{s} = 7$ and 8~TeV data, the CMS collaboration sets the upper bound as BR$(t\to cg) < 4.1\times 10^{-4}$~\cite{Khachatryan:2016sib}. On the other hand, utilizing 36 fb$^{-1}$ data at $\sqrt{s} = 13$~TeV and searching for $t\bar{t}$ events with one top decaying through the dominant SM channel $t\to bW$ and the other through the FCNC $t\to qZ~(q = u,c)$ decay, the ATLAS Collaboration puts an upper bound on the $t\to cZ$ channel to be BR$(t\to cZ) < 2.3 \times 10^{-4}$~\cite{ATLAS-CONF-2017-070}. Also, a bound of BR$(t\to cZ) < 4.9\times 10^{-4}$ is obtained by the CMS Collaboration using 19.7 fb$^{-1}$ data at $\sqrt{s} = 8$~TeV in the search for the associated production of top quark and $Z$ boson~\cite{Sirunyan:2017kkr}. Evidently, current and even future experiments would hardly be able to reach the sensitivity to probe decay widths of the orders predicted by the mUED and nmUED models. In comparison, a few BSM scenarios such as 2HDM~\cite{Eilam:1990zc, Abbas:2015cua, Gaitan:2017tka}, left-right symmetric model~\cite{Gaitan:2004by}, MSSM~\cite{Cao:2007dk}, $R$-parity violating SUSY~\cite{Bardhan:2016txk}, warped extra dimensional models~\cite{Agashe:2006wa, Gao:2013fxa}, and composite Higgs model~\cite{Agashe:2009di}, predict an enhanced branching ratio of these rare modes up to the level that can be probed in the future colliders.

\section{Summary and Conclusions} 
\label{sec:concl}

As an extension of our previous work~\cite{Dey:2016cve}, we perform a complete one-loop calculation of flavor-changing top quark decays in the $t\to cg$ and $t\to cZ$ channels in the context of minimal UED (mUED) and non-minimal UED (nmUED) scenarios. In the process of evaluating the results for the (n)mUED scenarios, we have also verified the SM estimates of the branching ratios given in the existing literature. 

We find that the decay widths, and hence the branching ratios, of $t\to cg$ and $t\to cZ$ do not alter much from the SM values in the mUED for allowed values of the inverse compactification radius. In the nmUED, however, the decay widths can be enhanced for some choices of the boundary localized kinetic term (BLKT) parameters. For example, for $R_{\Phi}\geq 10$ and $R_{f} = 0.1$, the decay width of $t\to cg$ can be two to three orders of magnitude higher than what is expected from the SM. However, current precision data and the high resonance dilepton data from the LHC can severely constrain this parameter choice. Similarly, for the higher values of $R_{\Phi}$ and smaller values of $R_{f}$, there can be some enhancement in the branching ratio of $t\to cZ$ too. This choice of parameter is also disfavored by the experimental data. In the case of a universal BLKT parameter $R_{\Phi} = R_{f}$, there is no enhancement in either branching ratios. 

We note in passing that the parameter choices made in this study are consistent with the parameter space allowed by the down sector FCNC observables related to the $B_{s}\to \mu^{+}\mu^{-}$ and $B\to X_{s}\gamma$ decays~\cite{Datta:2015aka, Datta:2016flx}. However, to put robust bounds on the model parameters requires a dedicated and correlated study of the down sector FCNC observables, {\it e.g.}, $B_{d,s}-\bar{B}_{d,s}$ oscillations, $\epsilon_{K}$, $K_{0}-\bar{K}_{0}$ oscillation, and various $K$-decays.

Finally, going to the next possible extension of the present model setup can lead to higher branching ratios for the rare top decays. There are two immediately possible extensions. Firstly, one can take non-universal and flavor-dependent BLKT parameters that can readily lead to top FCNC decays. In such a scenario, one must take into account the constraints coming from other flavor observables. Secondly, the KK parity violation would lead to new interactions which shall enhance these rare decay widths, since in this case one of the particles running in the loop can very well be SM particles and thus reduce the propagator suppression. We shall take up these extensions in the context of rare top decays, considered in this paper and our earlier paper, in our future work.

\paragraph*{Acknowledgements\,:} 
The work of CWC was supported in part by the Ministry of Science and Technology (MOST) of Taiwan under Grant No. MOST 104-2628-M-002-014-MY4. UKD acknowledges the support from the MOST under Grant No. MOST 107-2811-M-002-011.

\appendix
\section*{Appendix} \label{s:appen}
Here we give a few overlap integrals that are relevant for our calculations:
\begin{subequations}
\label{eq:overlapints}
\begin{align}
\label{eq:ovint1a}
I_{A}^{lk} &= \int_{0}^{\pi R}dy \left[
              1+r_{f}\{\delta(y)+\delta(y-\pi R)\}\right]
              f_{Qb_{L}}^{(l)}(y)
              ~f_{\phi}^{(k)}(y)(f_{W_{\mu}}^{(k)}(y))~f_{t_{R}}^{(0)}(y), \\
\label{eq:ovint2a}
I_{B}^{lk} &= \int_{0}^{\pi R}dy ~f_{Qb_{R}}^{(l)}(y) 
              ~f_{W_{5}}^{(k)}(y)~f_{t_{L}}^{(0)}(y),\\
\label{eq:ovint3a}
I_{C}^{k} &= \int_{0}^{\pi R}dy \left[
              1+r_{f}\{\delta(y)+\delta(y-\pi R)\}\right]
              f_{Qt_{L}}^{(0)}(y)~f_{\phi}^{(k)}(y)(f_{W_{\mu}}^{(k)}(y))
              ~f_{b_{R}}^{(0)}(y).  
\end{align}
\end{subequations}
Utilizing the normalization factor of the zero modes, these overlap integrals can also be expressed as
%
\begin{align}
I_{A,B,C}^{lk} = \frac{1}{\sqrt{r_{f}+\pi R}}I_{a,b,c}^{lk}, 
\end{align}
where
\begin{subequations}
\label{eq:new_overlapints}
\begin{align}
\label{eq:new_ovint1a}
I_{a}^{lk} &= \int_{0}^{\pi R}dy \left[
              1+r_{f}\{\delta(y)+\delta(y-\pi R)\}\right]
              f_{Qb_{L}}^{(l)}(y)
              ~f_{\phi}^{(k)}(y)(f_{W_{\mu}}^{(k)}(y)),\\
\label{eq:new_ovint1b}
I_{b}^{lk} &= \int_{0}^{\pi R}dy ~f_{Qb_{R}}^{(l)}(y) 
              ~f_{W_{5}}^{(k)}(y),\\
\label{eq:new_ovint1c}
I_{c}^{k} &= \frac{1}{\sqrt{r_{f}+\pi R}}\int_{0}^{\pi R}dy \left[
              1+r_{f}\{\delta(y)+\delta(y-\pi R)\}\right]
              f_{\phi}^{(k)}(y)(f_{W_{\mu}}^{(k)}(y)).
\end{align}
\end{subequations}
Below we enlist some of the relevant Feynman rules. All the momenta of the fields are assumed to be incoming. To avoid cluttering, we write $I_{c}$ in place of $I_{c}^{m}$ and $I_{a,b}$ in place of $I_{a,b}^{nm}$. Also $\beta = \frac{\pi + R_{\Phi}}{\pi + R_{f}}$. Obviously in mUED, the BLKT parameters are vanishing and the overlap integrals and $\beta$ will be unity, $\alpha_{n} = \frac12 \tan^{-1} \left(\frac{m_{j}}{n/R}\right)$ and $M_{\Phi k} = k/R$. As mentioned earlier, for mUED, the conservation of KK number will ensure the absence of (0)-(0)-($n$) couplings.

\fbox{
  \includegraphics[scale=0.55]{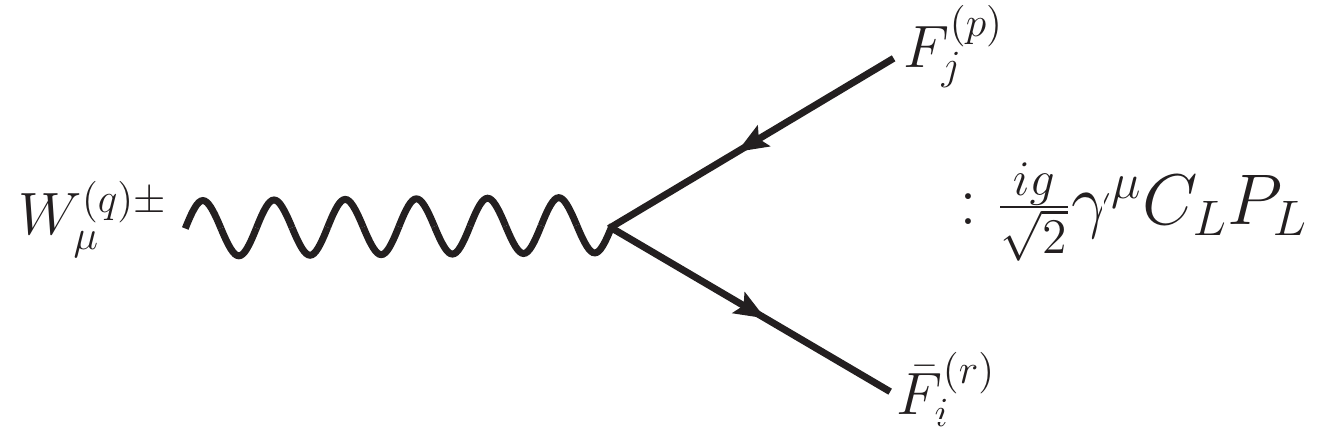}
}
\begin{flalign*}
&W^{(m)+}\bar{t}_{i}^{(0)}Q_{j}^{\prime(n)}  \colon   C_{L} = -I_{a}\sqrt \beta\cos\alpha_{n} V_{ij},
                                       ~~~~~~~~~ W^{(m)-}\bar{Q}_{j}^{\prime(n)}t_{i}^{(0)}  \colon  
                                      C_{L} = -I_{a}\sqrt \beta\cos\alpha_{n}V_{ij}^{*},&\\
&W^{(m)+}\bar{t}_{i}^{(0)}D_{j}^{\prime(n)}  \colon   C_{L} = I_{a}\sqrt \beta\sin\alpha_{n}V_{ij},
                                       ~~~~~~~~~~~~ 
W^{(m)-}\bar{D}_{j}^{\prime(n)}t_{i}^{(0)}  \colon  
                                      C_{L} = I_{a}\sqrt \beta\sin\alpha_{n}V_{ij}^{*},&\\
&W^{(m)+}\bar{t}_{i}^{(0)}b_{j}^{(0)}  \colon   C_{L} = I_{c}\sqrt {\beta}V_{ij},
                                      ~~~~~~~~~~~~~~~~~~~~~~~ 
W^{(m)-}\bar{b}_{j}^{(0)}t_{i}^{(0)}  \colon  
                                     C_{L} = I_{c}\sqrt{\beta} V_{ij}^{*}.&
\end{flalign*}

\fbox{
  \includegraphics[scale=0.55]{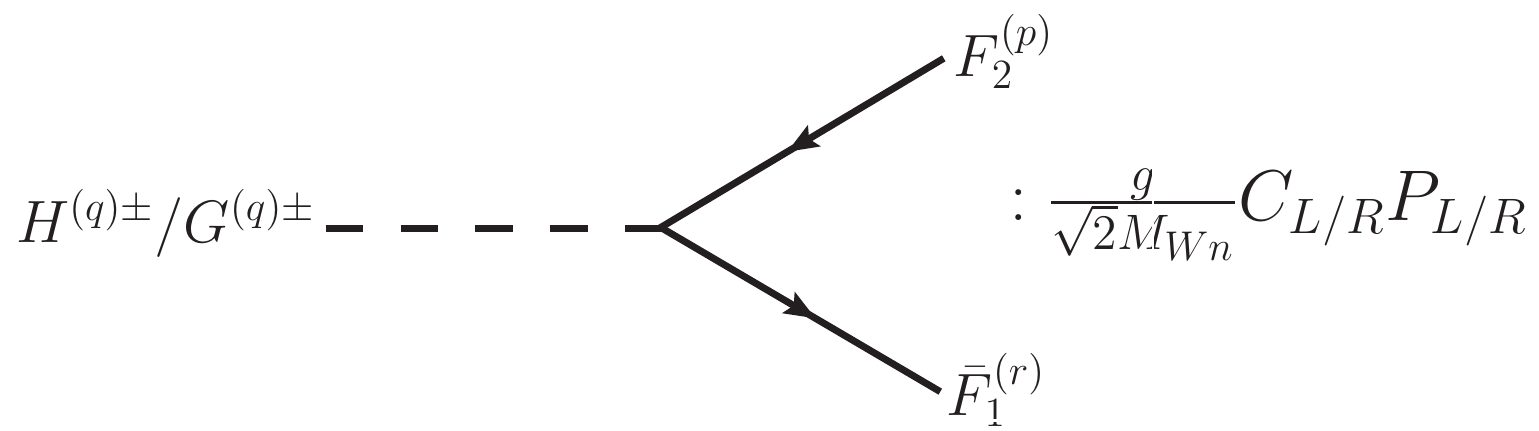}
}
\begin{flalign*}
&\bar{t}_{i}^{(0)}G^{(m)+}Q_{j}^{\prime(n)}  \colon  \left \lbrace \begin{aligned}
                                       C_{L} &= \sqrt \beta I_{a}m_{i}\cos\alpha_{n} V_{ij} \\
                                       C_{R} &= -\sqrt \beta\left(I_{a}m_{j}\sin\alpha_{n} + I_{b}M_{\Phi m}\cos\alpha_{n}\right)V_{ij}
                                      \end{aligned} \right.,&\\
&\bar{Q}_{j}^{\prime(n)}G^{(m)-}t_{i}^{(0)}  \colon  \left \lbrace \begin{aligned}
                                      C_{L} &=  \sqrt \beta\left(I_{a}m_{j}\sin\alpha_{n} + I_{b}M_{\Phi m}\cos\alpha_{n}\right)V_{ij}^{*}\\
                                      C_{R} &= -\sqrt \beta I_{a}m_{i}\cos\alpha_{n} V_{ij}^{*}
                                     \end{aligned} \right.,&\\
&\bar{t}_{i}^{(0)}G^{(m)+}D_{j}^{\prime(n)}  \colon  \left\lbrace \begin{aligned}
                                      C_{L} &= -\sqrt \beta I_{a}m_{i}\sin\alpha_{n} V_{ij} \\
                                      C_{R} &= \sqrt \beta\left(I_{a}m_{j}\cos\alpha_{n} - I_{b}M_{\Phi m}\sin\alpha_{n}\right)V_{ij}
                                      \end{aligned} \right.,&\\
&\bar{D}_{j}^{\prime(n)}G^{(m)-}t_{i}^{(0)}  \colon  \left\lbrace \begin{aligned}
                                      C_{L} &= -\sqrt \beta\left(I_{a}m_{j}\cos\alpha_{n} - I_{b}M_{\Phi m}\sin\alpha_{n}\right)V_{ij}^{*}\\
                                      C_{R} &= \sqrt \beta I_{a}m_{i}\sin\alpha_{n} V_{ij}^{*}
                                     \end{aligned} \right.,&\\
&\bar{t}_{i}^{(0)}H^{(m)+}Q_{j}^{\prime(n)}  \colon  \left \lbrace \begin{aligned}
                                      C_{L} &= -i\sqrt \beta I_{a}\frac{m_{i}M_{\Phi m}}{M_{W}}\cos\alpha_{n} V_{ij} \\
                                      C_{R} &= i\sqrt \beta\left(I_{a}\frac{m_{j}M_{\Phi m}}{M_{W}}\sin\alpha_{n} - I_{b}M_{W}\cos\alpha_{n}\right)V_{ij}
                                      \end{aligned} \right.,&\\
&\bar{Q}_{j}^{\prime(n)}H^{(m)-}t_{i}^{(0)}  \colon  \left \lbrace \begin{aligned}
                                      C_{L} &=  i\sqrt \beta\left(I_{a}\frac{m_{j}M_{\Phi m}}{M_{W}}\sin\alpha_{n} - I_{b}M_{W}\cos\alpha_{n}\right)V_{ij}^{*}\\
                                      C_{R} &= -i\sqrt \beta I_{a}\frac{m_{i}M_{\Phi m}}{M_{W}}\cos\alpha_{n} V_{ij}^{*}
                                     \end{aligned} \right.,&\\
&\bar{t}_{i}^{(0)}H^{(m)+}D_{j}^{\prime(n)}  \colon  \left\lbrace \begin{aligned}
                                      C_{L} &=  i\sqrt \beta I_{a}\frac{m_{i}M_{\Phi m}}{M_{W}}\sin\alpha_{n}V_{ij} \\
                                      C_{R} &= -i\sqrt \beta\left(I_{a}\frac{m_{j}M_{\Phi m}}{M_{W}}\cos\alpha_{n} + I_{b}M_{W}\sin\alpha_{n}\right)V_{ij}
                                      \end{aligned} \right.,&\\  
&\bar{D}_{j}^{\prime(n)}G^{(m)-}t_{i}^{(0)}  \colon  \left\lbrace \begin{aligned}
                                      C_{L} &= -i\sqrt \beta\left(I_{a}\frac{m_{j}M_{\Phi m}}{M_{W}}\cos\alpha_{n} + I_{b}M_{W}\sin\alpha_{n}\right)V_{ij}^{*}\\
                                      C_{R} &= i\sqrt \beta I_{a}\frac{m_{i}M_{\Phi m}}{M_{W}}\sin\alpha_{n} V_{ij}^{*}
                                     \end{aligned} \right.,&\\
&\bar{t}_{i}^{(0)}G^{(m)+}b_{j}^{(0)}  \colon  \left\lbrace \begin{aligned}
                                      C_{L} &=  -\sqrt{\beta} I_{c}m_{i}V_{ij} \\
                                      C_{R} &= \sqrt{\beta} I_{c} m_{j} V_{ij}
                                      \end{aligned} \right., ~~~~~~~~~~
\bar{b}_{j}^{(0)}G^{(m)-}t_{i}^{(0)}  \colon  \left\lbrace \begin{aligned}
                                      C_{L} &= -\sqrt{\beta} I_{c}m_{j}V_{ij}^{*}\\
                                      C_{R} &= \sqrt{\beta} I_{c} m_{i}V_{ij}^{*}
                                     \end{aligned} \right.,&\\
&\bar{t}_{i}^{(0)}H^{(m)+}b_{j}^{(0)}  \colon  \left\lbrace \begin{aligned}
                                      C_{L} &=  i \sqrt{\beta} I_{c} \frac{M_{\Phi m}}{M_{W}}m_{i}V_{ij} \\
                                      C_{R} &= -i \sqrt{\beta} I_{c} \frac{M_{\Phi m}}{M_{W}} m_{j} V_{ij}
                                      \end{aligned} \right., ~
\bar{b}_{j}^{(0)}H^{(m)-}t_{i}^{(0)}  \colon  \left\lbrace \begin{aligned}
                                      C_{L} &= -i \sqrt{\beta} I_{c} \frac{M_{\Phi m}}{M_{W}} m_{j}V_{ij}^{*}\\
                                      C_{R} &= i \sqrt{\beta} I_{c} \frac{M_{\Phi m}}{M_{W}}m_{i}V_{ij}^{*}
                                     \end{aligned} \right..&
\end{flalign*}

\fbox{
  \includegraphics[scale=0.6]{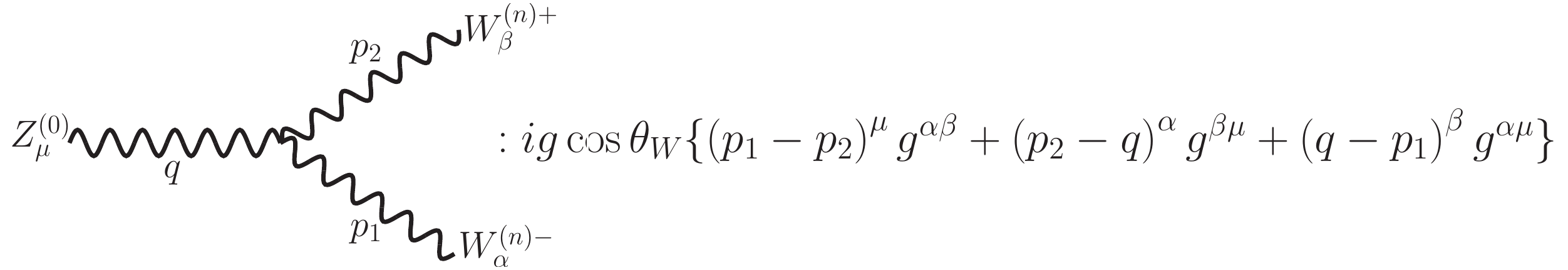}
}\\
\vfill
\fbox{
  \includegraphics[scale=0.6]{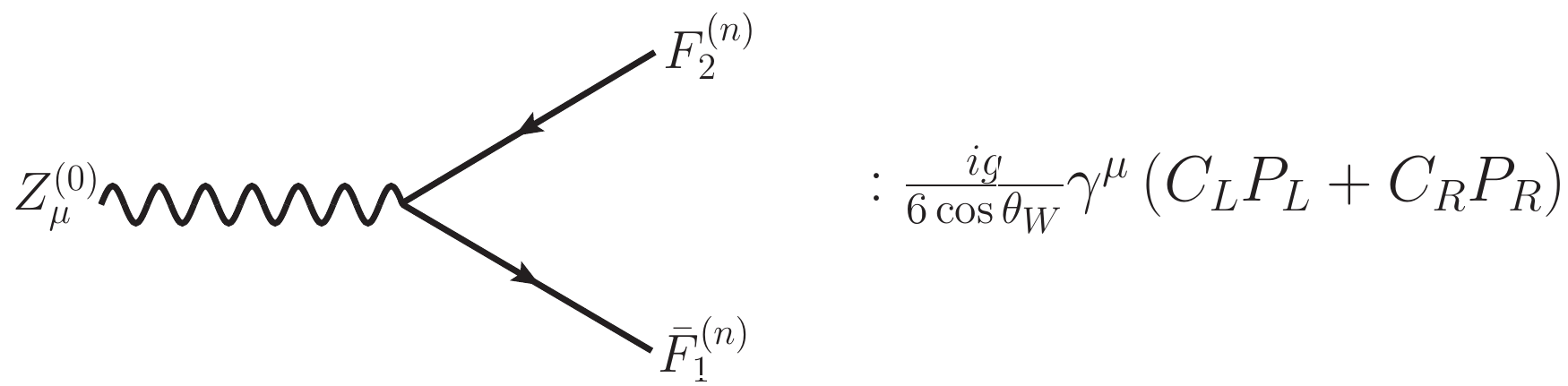}
}
\begin{flalign*}
&Z^{(0)}\bar{Q}_{t}^{\prime(n)}Q_{t}^{\prime(n)} \colon \left\lbrace \begin{aligned}
                                      C_{L} &= -4\sin^{2}\theta_{W}+3{\cos}^{2}\alpha_{n} \\
                                      C_{R} &= -4\sin^{2}\theta_{W}+3{\cos}^{2}\alpha_{n} \end{aligned}
                                      \right.,~
Z^{(0)}\bar{U}^{\prime(n)}U^{\prime(n)} \colon \left\lbrace \begin{aligned}
                                      C_{L} &= -4\sin^{2}\theta_{W}+3{\sin}^{2}\alpha_{n} \\
                                      C_{R} &= -4\sin^{2}\theta_{W}+3{\sin}^{2}\alpha_{n}
                                     \end{aligned} \right.,&\\
&Z^{(0)}\bar{Q}_{t}^{\prime(n)}U^{\prime(n)}  \colon  \left\lbrace \begin{aligned}
                                      C_{L} &= -3\sin\alpha_{n}\cos\alpha_{n} \\
                                      C_{R} &= -3\sin\alpha_{n}\cos\alpha_{n}
                                      \end{aligned} \right.,~~~~~~~~~ 
Z^{(0)}\bar{U}^{\prime(n)}Q_{t}^{\prime(n)}  \colon  \left\lbrace \begin{aligned}
                                      C_{L} &= -3\sin\alpha_{n}\cos\alpha_{n} \\
                                      C_{R} &= -3\sin\alpha_{n}\cos\alpha_{n}
                                      \end{aligned} \right..&
\end{flalign*}

\fbox{
  \includegraphics[scale=0.6]{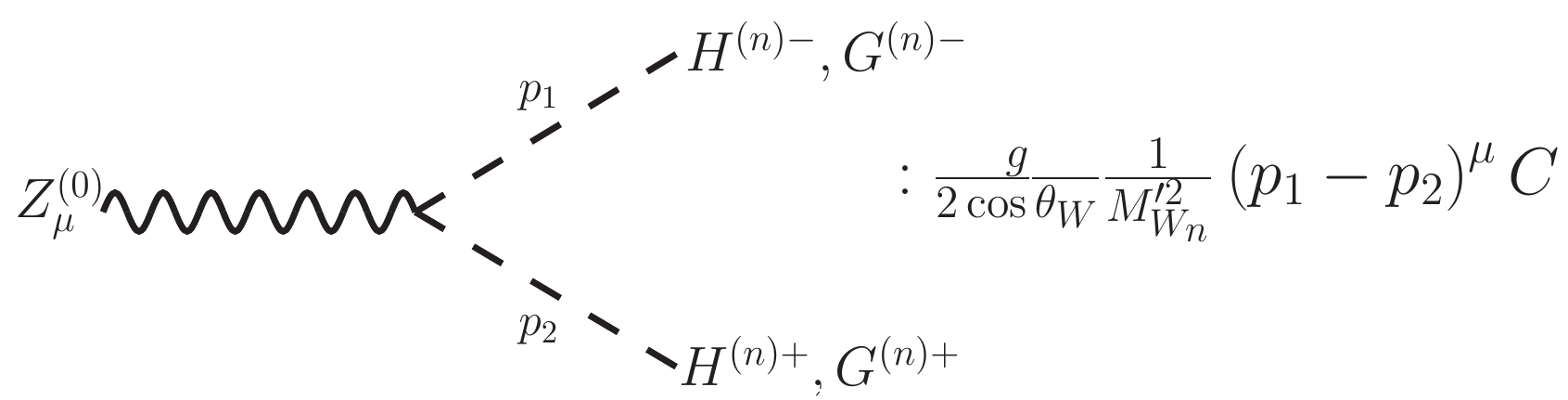}
}

\begin{flalign*}
&Z^{(0)}H^{(n)+}H^{(n)-}  \colon  C = i\{(-1+2\sin^{2}\theta_{W})M_{\Phi n}^{2} - 2\cos^{2}\theta_{W}M_{W}^{2}\},&\\
&Z^{(0)}G^{(n)+}G^{(n)-}  \colon  C = i\{(-1+2\sin^{2}\theta_{W})M_{W}^{2} - 2\cos^{2}\theta_{W}M_{\Phi n}^{2}\},&\\
&Z^{(0)}H^{(n)-}G^{(n)+}  \colon  C = -M_{\Phi n}M_{W},&\\
&Z^{(0)}G^{(n)-}H^{(n)+}  \colon  C = M_{\Phi n}M_{W}.&
\end{flalign*}

%
%
\fbox{
  \includegraphics[scale=0.6]{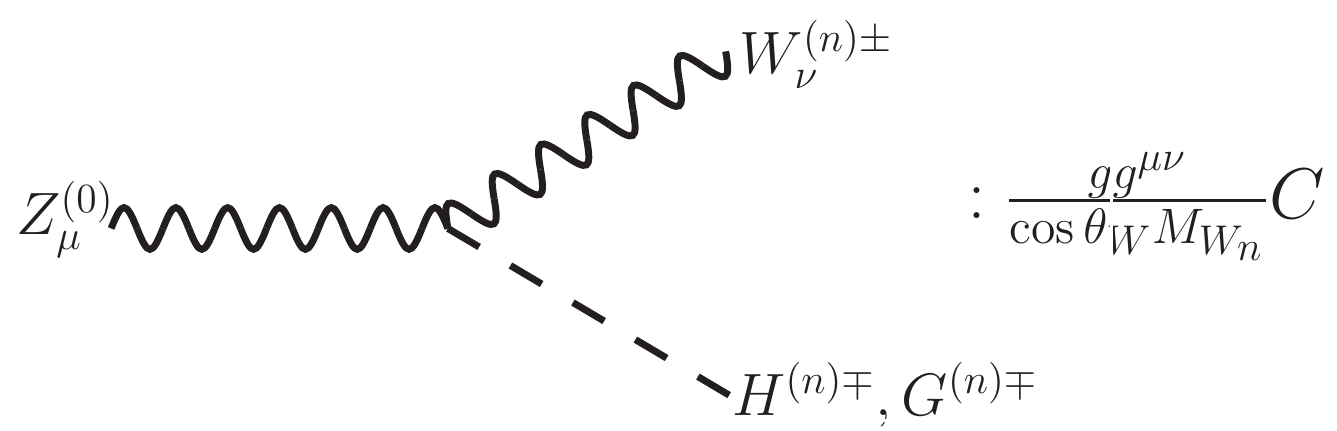}
}

\begin{flalign*}
&Z^{(0)}W^{(n)+}G^{(n)-}  \colon  C = \left(-M_{W}^{2}\sin^{2}\theta_{W} + M_{\Phi n}^{2}\cos^{2}\theta_{W}\right),&\\
&Z^{(0)}W^{(n)-}G^{(n)+}  \colon  C = \left(M_{W}^{2}\sin^{2}\theta_{W} - M_{\Phi n}^{2}\cos^{2}\theta_{W}\right),&\\
&Z^{(0)}W^{(n)+}H^{(n)-}  \colon  C = -iM_{\Phi n}M_{W},&\\
&Z^{(0)}W^{(n)-}H^{(n)+}  \colon  C = -iM_{\Phi n}M_{W}.&
\end{flalign*}

\fbox{
  \includegraphics[scale=0.6]{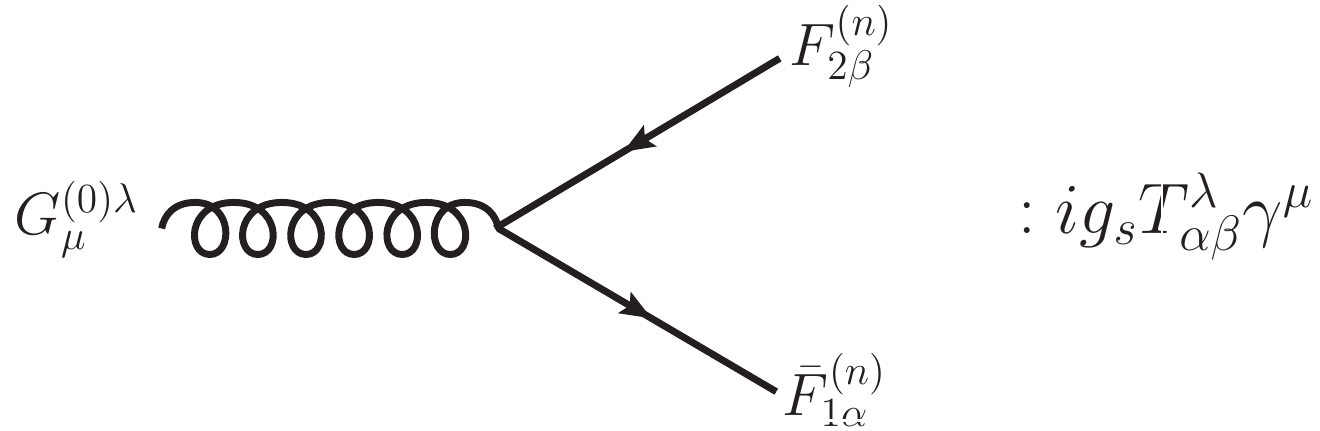}
}
%
%


\bibliographystyle{JHEP}
\bibliography{tcgtcZ_ref.bib}
\end{document}